\documentclass[a4paper,11pt]{article}

\usepackage{jheppub} 
\usepackage[T1]{fontenc}
\newcommand{\be}{\begin{equation}}
\newcommand{\ee}{\end{equation}}
\newcommand{\nn}{\nonumber}
\newcommand{\bea}{\begin{eqnarray}}
\newcommand{\eea}{\end{eqnarray}}
\newcommand{\bfig}{\begin{figure}}
\newcommand{\efig}{\end{figure}}
\newcommand{\bc}{\begin{center}}
\newcommand{\ec}{\end{center}}

\DeclareFontFamily{U}{wncy}{}
\DeclareFontShape{U}{wncy}{m}{n}{<->wncyr10}{}
\DeclareSymbolFont{mcy}{U}{wncy}{m}{n}
\DeclareMathSymbol{\sha}{\mathord}{mcy}{"58}

\title{Two-Loop integrals for CP-even heavy quarkonium production and decays}

\author[a,b]{Long-Bin Chen,}
\author[a]{Yi Liang,}
\author[a,b,c,1]{Cong-Feng Qiao\note{Corresponding author.}}

\affiliation[a]{School of Physics, University of Chinese Academy of Sciences,\\YuQuan Road 19A, Beijing 100049, China}
\affiliation[b]{CAS Center for Excellence in Particle Physics, Beijing 100049, China}
\affiliation[c]{Department of Physics \& Astronomy, York University, Toronto, ON M3J 1P3, Canada}

\emailAdd{chenlongbin@ucas.ac.cn}
\emailAdd{alavan@ucas.ac.cn}
\emailAdd{qiaocf@ucas.ac.cn}

\abstract{By employing the method of differential equations, we compute the various types of two-loop master integrals involved in CP-even heavy quarkonium exclusive production and decays. All the integrals presented in this paper can be casted into canonical forms and expressed in terms of Goncharov polylogarithms and Harmonic polylogarithms.  These master integrals are frequently used in the calculation of NNLO corrections of the heavy quarkonium production processes, as $\gamma^*\gamma\rightarrow Q\bar{Q}$, $e^+e^-\rightarrow \gamma+ Q\bar{Q}$,~and~$H/Z^0\rightarrow \gamma+ Q\bar{Q}$, and decay processes. They are also applicable in the calculation of NNLO corrections to CP-even quarkonium inclusive production and decay processes.}

\preprint{~}

\begin{document}
\maketitle
\flushbottom

\section{Introduction}
The production and decay mechanism of heavy quarkonium has been a longstanding topic since the discovery of heavy quarkonium \cite{ding,richter}. Large discrepancies between the experimental data and the leading order calculations stimulate  great amount of theoretical researches \cite{Abe:2002rb,Aubert:2005tj}. The complexity of the calculation for higher order QCD corrections of these processes made the investigation of this topic very challenging. Fortunately, the advance of Nonrelativistic Quantum Chromodynamics (NRQCD) factorization formalism enables people to study this mechanism more and more reliably \cite{NRQCD}. The progress in NRQCD calculations has deepen our understanding of strong interactions. It was found that the discrepancies can be remedied by introducing next-to-leading order (NLO) Quantum Chromodynamics (QCD) corrections \cite{Zhang:2005cha,Zhang:2006ay}. The calculations according to NRQCD recognize significant NLO corrections comparing to the LO results. However, people noticed that even after introducing the relatively large NLO corrections, the renormalization scale dependence and uncertainties can still be quite significant. Therefore, NNLO corrections will most likely be necessary if one hopes to further classify the issue.

The calculation of multi-loop corrections for heavy quarkonium processes is always considered to be difficult primarily due to the massive two-loop integrals it involves. In 1997, the first complete analytical calculation of the NNLO corrections to the processes of $J/\psi/\Upsilon \rightarrow e^+ e^-$ and $e^+ e^-\rightarrow J/\psi/\Upsilon$ was achieved \cite{beneke:1998,czarnecki:1998} (the NNNLO corrections to the same processes have not been obtained until 2014 \cite{Beneke:2014qea}). Since then, NNLO corrections to the heavy quarknoium processes have been studied extensively \cite{czarnecki:2001,bc22,Chen:2015csa}. Recently, the NNLO corrections to the process $\gamma\gamma^* \rightarrow \eta_c$ were calculated \cite{Feng:2015uha}. We notice that most of the master integrals computed in that work are in non-physical region. Moreover, the calculations of the integrals were performed purely numerically. The accuracy of the numerical approach is limited by the applicability of the numerical packages available. Higher accuracy is expected, especially for the processes in physical region. For example, the master integrals involved in the computation of $e^+e^- \rightarrow \gamma^* \rightarrow \gamma+\eta_c/\eta_b$ are all in physical region. Due to the limitation of today's numerical packages, a full numerical calculation of these integrals will most certainly be unreliable. Analytical calculations of the master integrals are highly desired.

Threshold expansion \cite{beneke:1997} is widely used in the calculations of loop-corrections for heavy quarkonium production and decay. This method is not only very convenient, but also provides an implicit definition of NRQCD in dimensional regularization \cite{beneke:1998}. Within threshold expansion, only contributions coming from the hard loop momenta are needed, the contributions from soft, potential, and ultrasoft loop momenta can all be factored out. Thanks to this method, the calculations of heavy quarkonium processes become much more efficient than the calculations of corresponding heavy quark processes.

A powerful method to evaluate the master integrals analytically is the method of differential equation \cite{Kotikov:1990kg, Kotikov:1991pm, Remiddi:1997ny, Gehrmann:1999as, Argeri:2007up}. Along with the recent development \cite{Henn:2013pwa,Henn:2013nsa,Henn:2014qga,Argeri:2014qva}, this method is becoming more and more powerful. It is pointed out  that for a generic multi-loop calculation, a suitable basis of master integrals can be chosen, so that the corresponding differential equations are simplified \cite{Henn:2013pwa}, and their iterative solution becomes straightforward in terms of dimensional regularization parameter $\epsilon=\frac{4-D}{2}$. Following this proposal, substantive analytical computations of various phenomenology processes have been completed \cite{Henn:2013woa,Henn:2014lfa,Gehrmann:2014bfa,Caola:2014lpa,DiVita:2014pza,Bell:2014zya,Huber:2015bva,Bonciani:2015eua,Gehrmann:2015dua,Grozin:2015kna,Bonciani:2016ypc}.

The Feynman diagrams of the NNLO corrections to CP-even heavy quarkonium exclusive production  $\gamma^*\gamma\rightarrow Q\bar{Q}$ and $e^+ e^-\rightarrow\gamma^* \rightarrow \gamma+ Q\bar{Q}$ share the same topology (both massless and massive "light-by-light" diagrams are included).  By performing reducitons of the amplitudes, the scalar integrals involved are reduced to 133 master integrals. We notice that 86 out of the 133 master integrals can be expressed in terms of Harmonic polylogarithms and Goncharov polylogarithms. In this work, we will focus on the analytical solutions of these integrals.

The paper is organized as follows. In section 2, we introduce the kinematics and notations for CP-even quarkonium exclusive production processes. We also present the generic form of the differential equations with respect to the kinematics variables. In section 3, the Goncharov polylogarithms as well as Harmonic polylogarithms are introduced.
In section 4, the canonical basis is explicitly presented, followed by the discussion of their solutions. In sections 5, the determination of the boundary conditions, as well as the analytical continuation are explained. Discussions and conclusions are made in section 6. All the analytical results up to weight four from our computation are collected in an ancillary file that we submit to the \textbf{arXiv}, the results up to weight three are listed in appendix.A.

\section{Notations and Kinematics}

\begin{figure}[t]
\begin{center}
\includegraphics[scale=0.56]{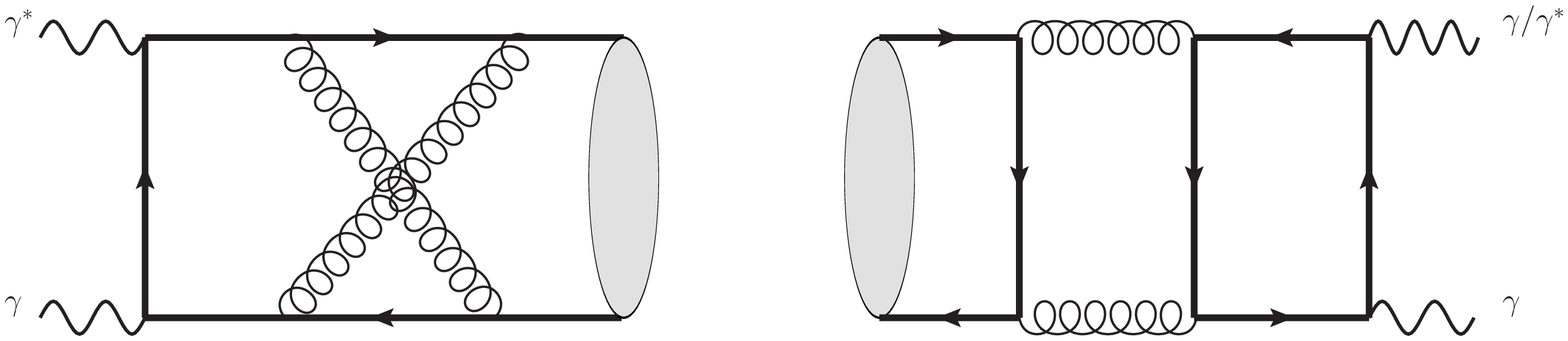}
\caption{Sample of Feynman diagrams contributing to the two-loop QCD corrections of $\gamma^*+ \gamma \rightarrow  Q \bar{Q}$ and  $Q \bar{Q} \rightarrow  \gamma^*/\gamma + \gamma$. }
\label{sample}
\end{center}
\end{figure}

We consider the kinematics for the production of heavy quarkonium through $\gamma^*\gamma$ collision, and the process associated with a photon via a virtual photon.
\bea
\gamma^*(k_1)+ \gamma (k_2) \rightarrow  Q(k_q) \bar{Q}(k_{\bar{q}}) ,\label{pro1} \\
\gamma^*(k_1)\rightarrow Q(k_q) \bar{Q}(k_{\bar{q}}) + \gamma (k_2),\label{pro2}
\eea
where $k_1^2 = 2 ss , k_2^2 = 0$ and $k_q^2 = k_{\bar{q}}^2 = m_q^2$. Sample of Feynman diagrams contributing to the NNLO corrections for CP-even heavy quarkonium exclusive production and decay are showed in Fig.(\ref{sample}). For processes (\ref{pro1})  in Euclidean region with $ss<0$,  we have the following relation
\be
(k_1+ k_2)^2 = (k_q + k_{\bar{q}})^2 = 4 m_q^2,
\ee
For processes (\ref{pro2}) in Minkowski region  with $2ss > 4 m_q^2$,  we have
\be
(k_1- k_2)^2 = (k_q + k_{\bar{q}})^2 = 4 m_q^2,
\ee
Within the threshold expansion approach, we take the momentum of quark and anti-quark to be equal $k_q = k_{\bar{q}}$.

In order to express the integrals more compactly, we introduce three dimensionless variables $x$, $y$ and $z$ as
\bea
\frac{ss}{m_q^2} = -\frac{(1-x)^2 }{2x}=(y+2)=(z+1) \, .\label{xyz}
\eea

The QCD corrections to the processes (\ref{pro1}) and (\ref{pro2}) are calculated using Feynman diagram approach. After some manipulations, the amplitudes can be expressed in terms of a set of scalar integrals. The calculation of these scalar integrals always turns out to be the most difficult part in the whole work. We first use packages $\textbf{FIRE}$ \cite{Smirnov:2008iw,Smirnov:2013dia,Smirnov:2014hma} to reduce the group of scalar integrals into a minimum set of independent master integrals. $\textbf{FIRE}$ is also adopted in the following derivations of differential equations.

The first step of deriving differential equations is taking derivatives of lorentz invariant, and writing them down as linear combinations of master integrals. The derivatives of the external momenta can be transformed to the derivatives of $ss$ and $m_q^2$
\bea
k_i\cdot\frac{\partial}{\partial k_j}=k_i\cdot\frac{\partial ss}{\partial k_j} + k_i\cdot\frac{\partial m_q^2}{\partial k_j} \, ,
\eea
where $i=1,2$. Using these linear equations, we can write the derivatives $\frac{\partial }{\partial ss}$ in terms of the derivatives $k_i\cdot\frac{\partial}{\partial k_j}$
\bea
2 ss \frac{\partial}{\partial ss} = k_1\cdot \frac{\partial}{\partial k_1} + \left(\frac{ss + 2 m_q^2}{ss - 2 m_q^2}\right) k_2\cdot \frac{\partial}{\partial k_2}\, .
\eea
The transform of the derivatives can readily be obtained according to (\ref{xyz}). With the variables chosen above, we express the integrals in terms of the so-called Goncharov polylogarithms and Harmonic polylogarithms, which will be discussed in the next section.

\section{Goncharov polylogarithms and Harmonic polylogarithms}

The Goncharov polylogarithms (GPLs) \cite{Goncharov:1998kja} are defined as follow
\bea
G_{a_1,a_2,\ldots,a_n}(x) &\equiv & \int_0^x \frac{\text{d} t}{t - a_1} G_{a_2,\ldots,a_n}(x)\, ,\\
G_{\overrightarrow{0}_n}(x) & \equiv & \frac{1}{n!}\log^n x\, .
\eea

They can be viewed as a special case belonging to a more general type of integrals called Chen-iterated integrals \cite{Chen}. If all the index $a_i$ belong to the set $\{0, \pm 1\}$, the Goncharov polylogarithms turns into the well-known Harmonic polylogarithms (HPLs) \cite{Remiddi:1999ew}
\bea
H_{\overrightarrow{0}_n}(x) &=&G_{\overrightarrow{0}_n}(x)\, ,\\
H_{a_1,a_2,\ldots,a_n}(x) &=&(-1)^k G_{a_1,a_2,\ldots,a_n}(x),
\eea
where $k$ equals to the times of element $(+1)$ taken in $(a_1,a_2,\ldots,a_n)$\, .
The GPLs fulfil the following shuffle rules
\bea
G_{a_1,\ldots,a_m}(x)G_{b_1,\ldots,b_n}(x) &=& \sum_{c\in a \sha b} G_{c_1, c_2,\ldots,c_{m+n}}(x)\, .
\eea
Here, $a \sha b$ is composed of the shuffle products of list a and b. It is defined as the set of
the lists containing all the elements of a and b, with the ordering of the elements
of a and b preserved. The GPLs and HPLs can be numerically evaluated within the GINAC implementation \cite{Vollinga:2004sn,Bauer:2000cp}. A Mathematica package {\bf HPL} \cite{Maitre:2005uu,Maitre:2007kp} is available to reduce and evaluate the HPLs.  Both the GPLs and HPLs  can be transformed to the function of  $\ln, \text{Li}_n$ and $\text{Li}_{22}$  up to weight four, with the methods and packages described in \cite{Frellesvig:2016ske}.

\section{The canonical basis}

\begin{figure}[t]
\begin{center}
\includegraphics[scale=0.44]{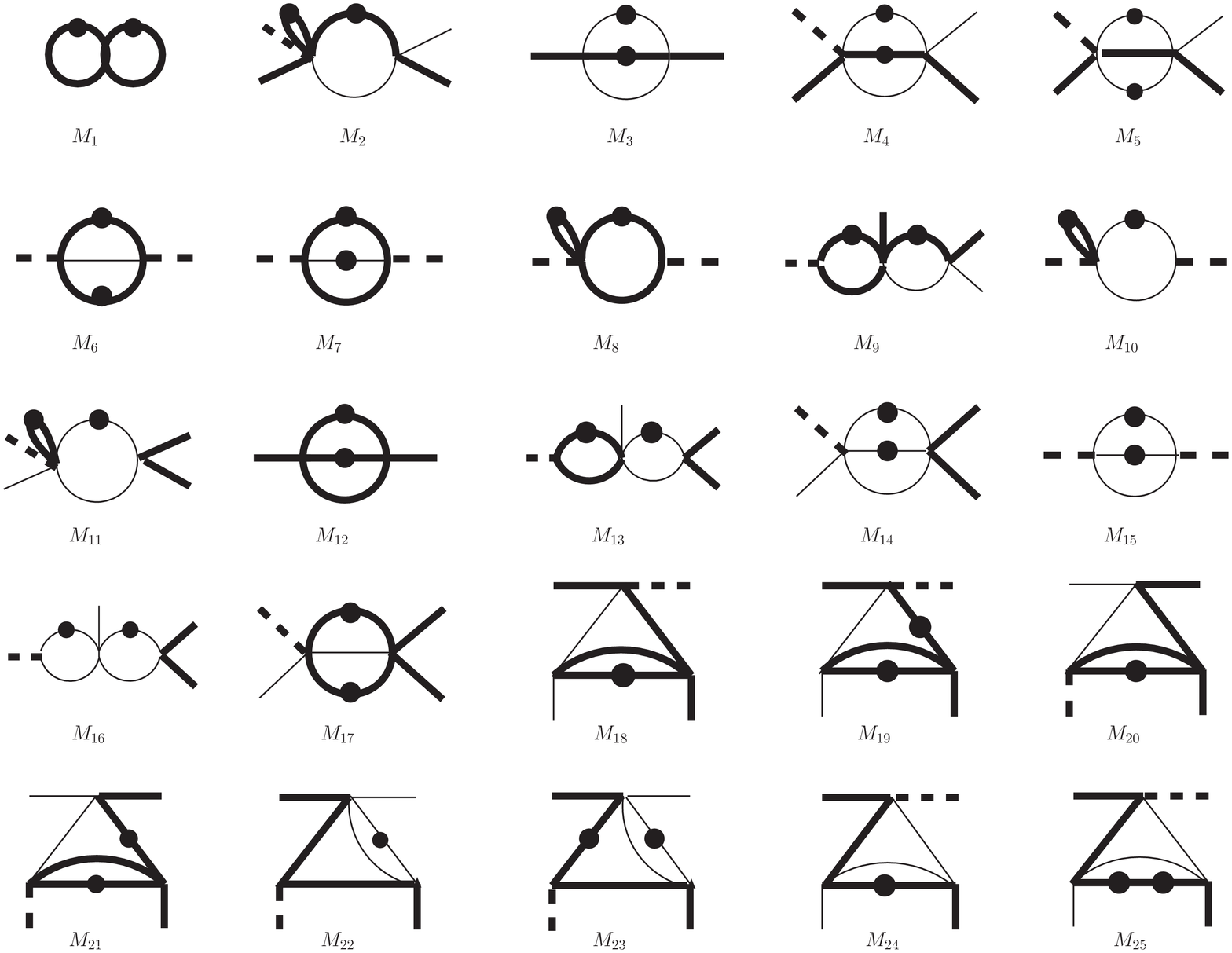}
\includegraphics[scale=0.44]{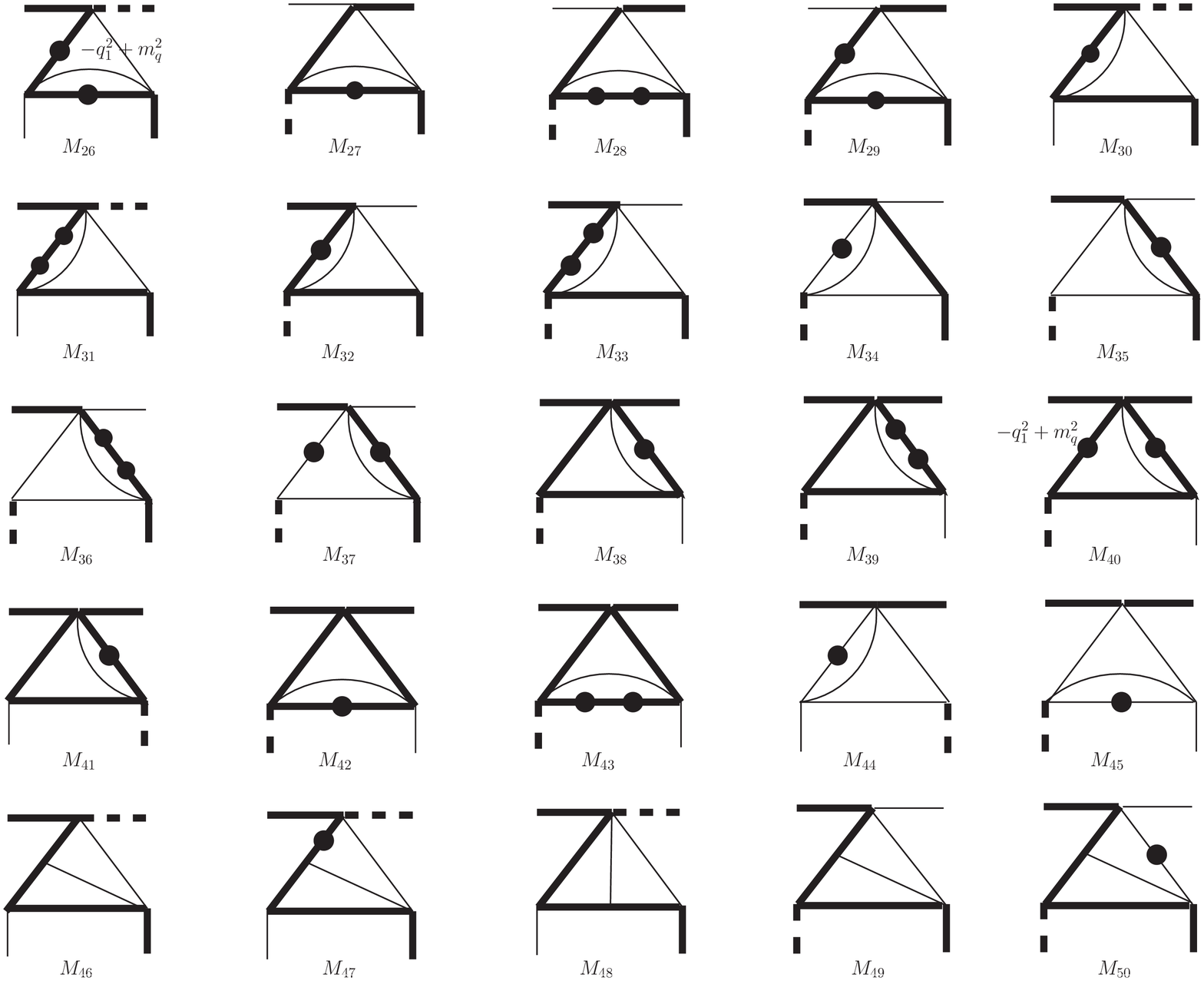}
\end{center}
\end{figure}
\begin{figure}[t]
\begin{center}
\includegraphics[scale=0.44]{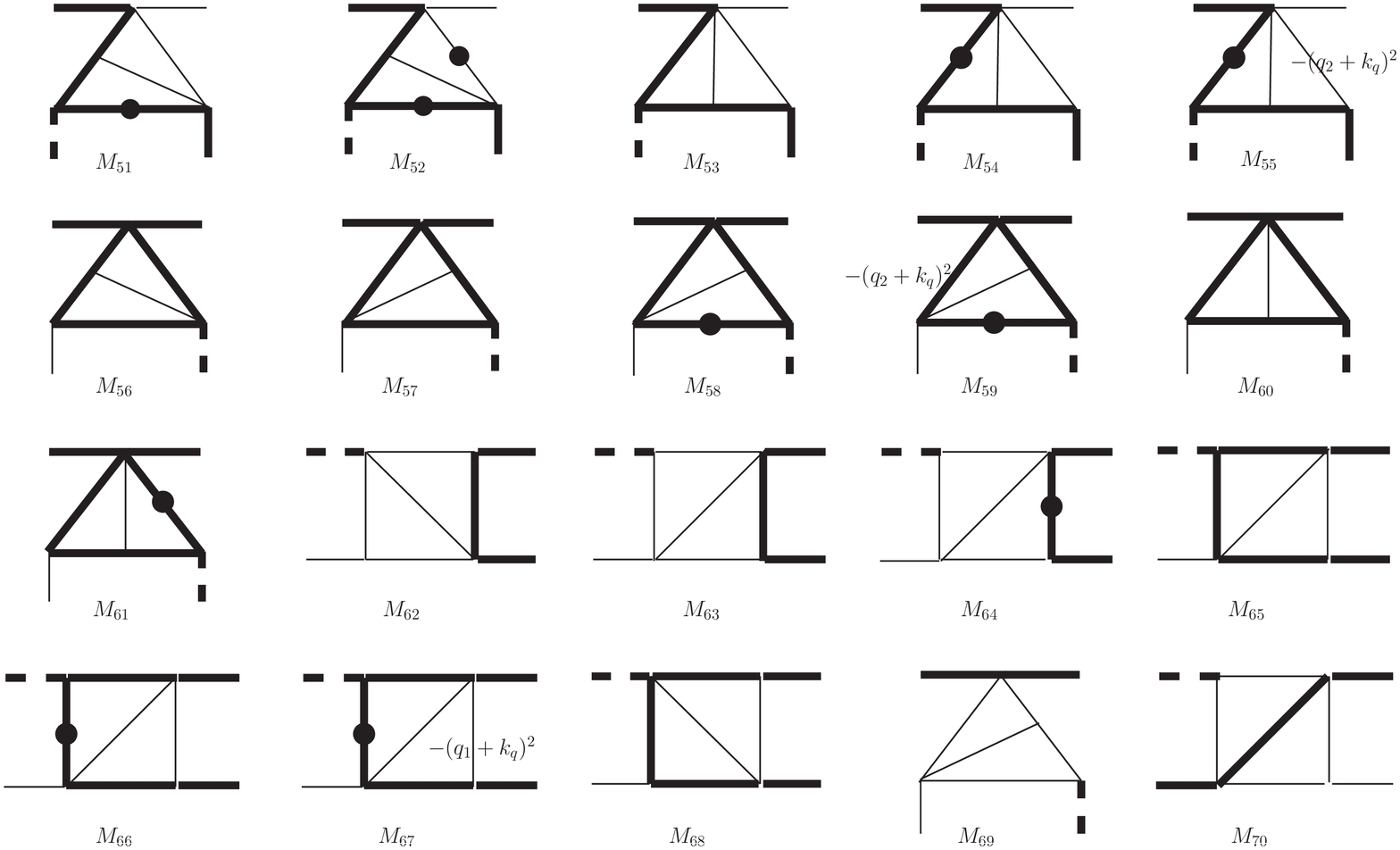}
\includegraphics[scale=0.44]{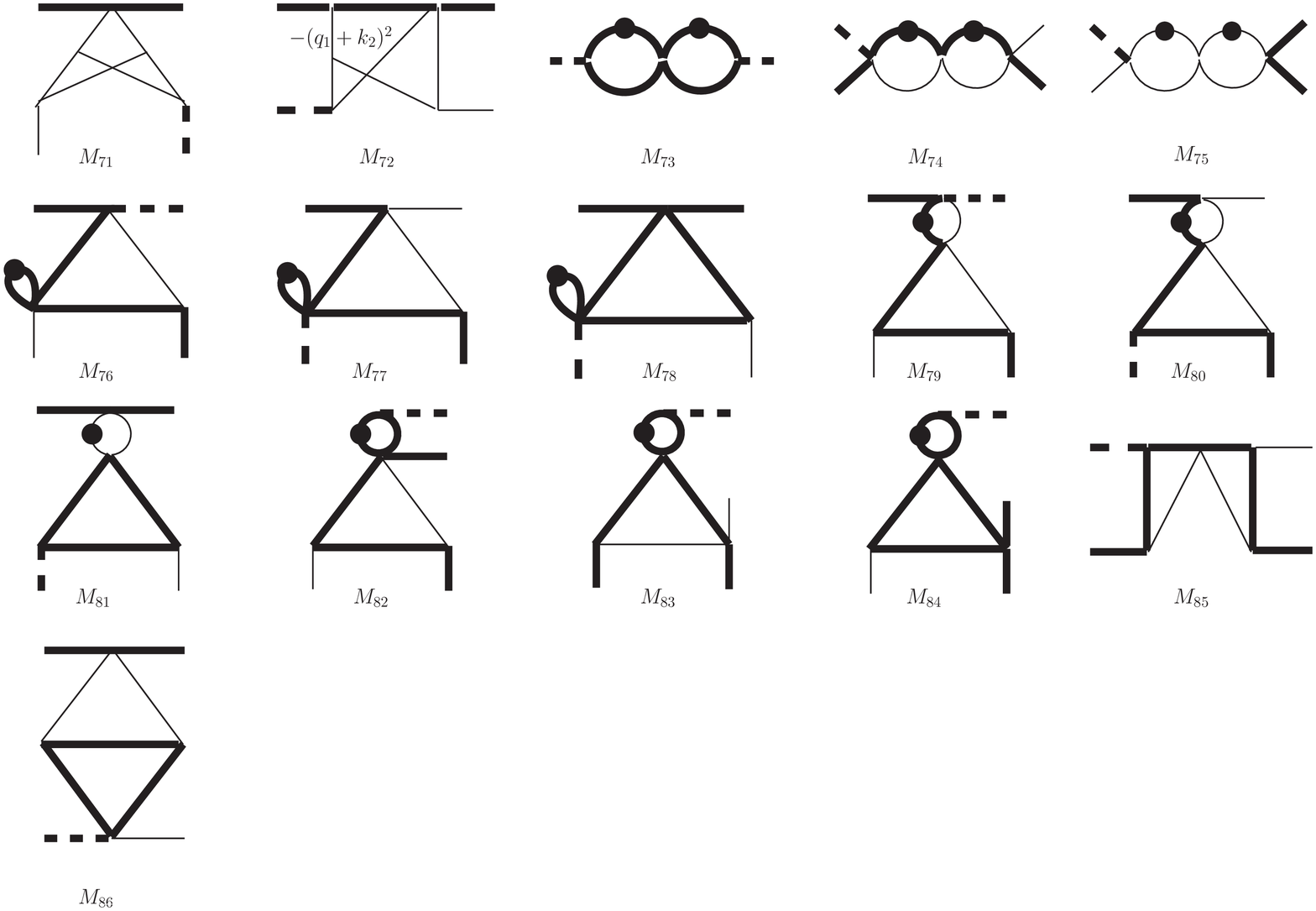}
\caption{Set of master integrals that can be cast into canonical form and expressed in terms of GPLs. The thin lines denote massless propagators and on-shell massless external particles; the thick lines denote massive quark propagators and on-shell external quarks. The dash lines denote off-shell external particles with squared momentum equal to $2 ss$. A dot on a propagator indicates that the power of the propagator is raised to 2. Two dots means that the propagator is raised to power 3.}
\label{midiag}
\end{center}
\end{figure}

It is widely known that the two-loop sunrise integrals with non-zero masses and off-shell external legs cannot be expressed only in terms of multiple polylogarithms. They can be evaluated and expressed, however, in terms of elliptic multiple polylogarithms \cite{Adams:2015gva,Adams:2015ydq}. All the integrals of the processes that we concern can be reduced to a minimum set of 133 master integrals. 41 of these master integrals cannot be cast into canonical form and then expressed in terms of GPLs. Among the 41 integrals, the solutions of 2 full massive sunrise integrals with off-shell external legs and 1 kite integral  have been discussed \cite{Adams:2015gva,Adams:2015ydq,Adams:2014vja,Remiddi:2013joa,Adams:2016xah,Remiddi:2016gno}. 6 of the 133 master integrals were calculated in terms of Chen-iterated integrals (integrals $f_{13}^B,f_{14}^B,f_{(29,\ldots,32)}^B$ in \cite{Bonciani:2016qxi}). Here we focus on the remaining 86 integrals. They all can be cast into canonical form and expressed in terms of HPLs and GPLs. The remaining 41 integrals  will be discussed in detail elsewhere \cite{inpreparation}.

Fig.\ref{midiag} shows the 86 master integrals we are discussing in this paper. To the best of our knowledge, master integrals $M_{18\ldots37}$, $M_{46\ldots55}$ ,$M_{62\ldots68}$, $M_{70}$ and $M_{72}$ have not been given in any literature before.

The corresponding of 86 linear differential equations can be expressed, via a suitable basis choice of the master integrals,
into the canonical form \cite{Henn:2013pwa}
\bea
\text{d}~{\bf F} = \epsilon \, (\text{d}~{\bf A}) \, {\bf F} \, .
\label{sysdeq}
\eea
Here ${\bf F}$ is a vector of 86 canonical master integrals. The $\epsilon$ dependence is then completely factorized from the
matrix $\text{d} {\bf A}$. The vector ${\bf F}$ depends only on the dimensionless variables $\{ x, y, z\}$ , defined in Eq.~(\ref{xyz}). The matrix  $\text{d} {\bf A}$ represents a total differential as follow
\bea
\text{d}~{\bf A}=\sum_{k=1}^{9}{\bf \text{R}}_k~\text{d}~\log(l_k)\, ,
\eea
where ${\bf \text{R}}_k$ are constant matrices that contain only rational numbers, and the $l_k$ are linear functions of $\{ x, y, z\}$. The so-called alphabet of the GPLs is preserved in order to obtain the result
\be
l_k \in \{x_n, x_n-1, x_n+1, x_n-i, x_n+i, x_n+\frac{1 - \sqrt{3}i}{2}, x_n+\frac{1 + \sqrt{3}i}{2}, x_n+\frac{1}{3},x_n+3\} \label{ind}\, ,
\ee
where $x_n\in\{ x, y, z\}$ . The constant matrices $R_k$ with rational numbers are collected in an ancillary file called "Matrix.txt" that we submit to the \textbf{arXiv}.

The search of the canonical basis is based on an experimental approach with trial and error, with the aid of Mathematica code written by ourselves. We notice that packages for choosing a canonical basis appeared recently \cite{Prausa:2017ltv,Gituliar:2017vzm}.

The vector of basis ${\bf F}$ is built up with 86 functions $F_i(ss,m_q,\epsilon)$, defined in terms of the master integrals $M_i$ drawn in Fig.(\ref{midiag}),
\bea
F_1 & = & \, \epsilon^2 \, M_1 \, , \label{f1}\\
F_2 & = & \, \epsilon^2 \, (ss - m_q^2) \,  M_{2} \, , \label{f2}\\
F_3 & = & \, \epsilon^2 \,  m_q^2\,   M_3\, ,\label{af3}\\
F_4 & = & \, \epsilon^2 \, (ss - m_q^2) \,  M_4 \, , \label{f4}\\
F_5 & = & \, \epsilon^2 \, \left[ (ss - 2 m_q^2) \,  M_5 - 2 m_q^2  M_4 \right] \, ,\label{f5}\\
F_6 & = & \, \epsilon^2 \, ss  M_6 \, , \label{f6}\\
F_7 & = & \, \epsilon^2 \, \sqrt{ ss}\sqrt{ss - 2 m_q^2}( M_6 + 2 M_7)/2 \, , \label{f7}\\
F_8 & = & \, \epsilon^2 \, \sqrt{ ss}\sqrt{ss - 2 m_q^2} \, M_{8} \, , \label{f8}\\
F_9 & = & \, \epsilon^2 \, (ss - m_q^2) \sqrt{ ss}\sqrt{ss - 2 m_q^2} \, M_{9} \, , \label{f9}\\
F_{10} & = & \, \epsilon^2 \, ss\, M_{10} \, , \label{f10}\\
F_{11} & = & \, \epsilon^2 \, 2 m_q^2\, M_{11} \, , \label{f11}\\
F_{12} & = & \, \epsilon^2 \, m_q^2\, M_{12} \, , \label{f12}\\
F_{13} & = & \, \epsilon^2 \, m_q^2\, \sqrt{ ss}\sqrt{ss - 2 m_q^2} \,  M_{13}  \, , \label{f13}\\
F_{14} & = & \, \epsilon^2 \, m_q^2\, M_{14} \, , \label{f14}\\
F_{15} & = & \, \epsilon^2 \, ss \, M_{15} \, , \label{f15}\\
F_{16} & = & \, \epsilon^2 \, m_q^2\, ss \, M_{16} \, , \label{f16}\\
F_{17} & = & \, \epsilon^2 \, 2\, m_q^2\, M_{17} \, , \label{f17}\\
F_{18} & = & \, \epsilon^3 \, (ss - 2 m_q^2) \,  M_{18}      \, , \label{m413a1}\\
F_{19} & = & \, \epsilon^2 \, \left[ ss\, (ss - 2 m_q^2) \,  M_{19} + 3 ss/2 \, M_{12} \right]
\, , \label{m413a2}\\
F_{20} & = & \, \epsilon^3 \, (ss - 2 m_q^2) \,  M_{20}   \, , \label{m414a1}\\
F_{21} & = & \, \epsilon^2 \,  \left[ (ss - 2 m_q^2)^2 \,  M_{21} + (ss - 2 m_q^2)/2\left( M_6 + 2 M_7 \right) \right]   \, , \label{m414a2}\\
F_{22} & = & \, \epsilon^3 \,  (ss - 2 m_q^2) \,  M_{22}  \, , \label{m416a1}\\
F_{23} & = & \, \epsilon^2 \, \sqrt{ ss}\sqrt{ss - 2 m_q^2}\left[ (ss - 2 m_q^2) \,  M_{23} + 3 \, \epsilon\, M_{22} \right]
\, , \label{m416a2}\\
F_{24} & = & \, \epsilon^3 \, (ss - 2 m_q^2) \,  M_{24}   \, , \label{m417a1}\\
F_{25} & = & \, \epsilon^2 \, m_q^2 (ss - 2 m_q^2) \,  M_{25}  \, , \label{m417a2}\\
F_{26} & = & \, \epsilon^2 \, \frac{1}{2(ss-2 m_q^2)}\left[2(ss - m_q^2)((ss - 3 m_q^2)M_{26} -2 m_q^2(ss - 2 m_q^2)M_{25} \right. \nn\\
& & + \left. 2 \epsilon m_q^2 M_{24}) + 4 m_q^4 M_3 -m_q^2 M_1  \right]  \, , \label{m417a3}\\
F_{27} & = & \, \epsilon^3 \, (ss - 2 m_q^2) \,  M_{27}  \, , \label{m418a1}\\
F_{28} & = & \, \epsilon^2 \, m_q^2\, (ss - 2 m_q^2) \,  M_{28}  \, , \label{m418a2}\\
F_{29} & = & \, \epsilon^2 \, (ss - m_q^2)((2 ss- 3 m_q^2) M_{29} - 2 m_q^2 M_{28} + 3 \epsilon M_{27})  \, , \label{m418a3}\\
F_{30} & = & \, \epsilon^3 \, (ss - 2 m_q^2) \,  M_{30}  \, , \label{m419a1}\\
F_{31} & = & \, \epsilon^2 \,  m_q^2 (ss - 2 m_q^2) \,  M_{31}  \, , \label{m419a2}\\
F_{32} & = & \, \epsilon^3 \, (ss - 2 m_q^2) \,  M_{32}  \, , \label{m420a1}
\eea
\bea
F_{33} & = & \, \epsilon^2 \,  m_q^2 (ss - 2 m_q^2) \,  M_{33}  \, , \label{m420a2}\\
F_{34} & = & \, \epsilon^3 \, (ss - 2 m_q^2) \,  M_{34}  \, , \label{m422}\\
F_{35} & = & \, \epsilon^3 \, (ss - 2 m_q^2) \,  M_{35}  \, , \label{m424a1}\\
F_{36} & = & \, \epsilon^2 \, m_q^2 (ss - 2 m_q^2) \,  M_{36}  \, , \label{m424a2}\\
F_{37} & = & \, \epsilon^2 \, (ss - m_q^2) (ss M_{37}- 2 m_q^2 M_{36} + 3 \epsilon M_{35})   \, , \label{m424a3}\\
F_{38} & = & \, \epsilon^3 \, (ss - 2 m_q^2) \,  M_{38}  \, , \label{m425a1}\\
F_{39} & = & \, \epsilon^2 \, m_q^2 (ss - 2 m_q^2) \,  M_{39}  \, , \label{m425a2}\\
F_{40} & = & \, \epsilon^2 \, \left[\frac{\sqrt{ss}(3 ss - 8 m_q^2)}{\sqrt{ss - 2 m_q^2}} \,  M_{40}
+ 2  m_q^2(ss - 2 m_q^2)(1 - \frac{2\sqrt{ss -2 m_q^2}}{\sqrt{ss}})M_{39}  \right. \nn\\
& & + \left. 2 \epsilon (\frac{(3 ss -4 m_q^2)\sqrt{ss- 2 m_q^2}}{\sqrt{ss}}-3(ss - 2 m_q^2)) M_{38} \right. \nn\\
& & + \left. \frac{2 m_q^2}{\sqrt{ ss}\sqrt{ss - 2 m_q^2}}M_{17}\right] \, , \label{m425a3}\\
F_{41} & = & \, \epsilon^3 \, (ss - 2 m_q^2) \,  M_{41}  \, , \label{m426}\\
F_{42} & = & \, \epsilon^3 \, (ss - 2 m_q^2) \,  M_{42}  \, , \label{m427a1}\\
F_{43} & = & \, \epsilon^2 \, m_q^2(ss - 2 m_q^2) \,  M_{43}  \, , \label{m427a1}\\
F_{44} & = & \, \epsilon^3 \, (ss - 2 m_q^2) \,  M_{44}  \, , \label{m429}\\
F_{45} & = & \, \epsilon^3 \, (ss - 2 m_q^2) \,  M_{45}  \, , \label{m431}\\
F_{46} & = & \, \epsilon^4 \, (ss - 2 m_q^2) \,  M_{46}  \, , \label{m59a1}\\
F_{47} & = & \, \epsilon^3 \,  m_q^2 (ss - 2 m_q^2) \,  M_{47}  \, , \label{m59a2}\\
F_{48} & = & \, \epsilon^4 \, (ss - 2 m_q^2) \,  M_{48}  \, , \label{m511}\\
F_{49} & = & \, \epsilon^4 \, (ss - 2 m_q^2) \,  M_{49}  \, , \label{m510a1}\\
F_{50} & = & \, \epsilon^3 \, (ss - 2 m_q^2)^2 \,  M_{50}  \, , \label{m510a2}\\
F_{51} & = & \, \epsilon^3 \, \sqrt{ ss}\sqrt{ss - 2 m_q^2}(ss - 2 m_q^2) \,  M_{51}  \, , \label{m510a3}\\
F_{52} & = & \, \epsilon^2 \, \left[ m_q^2(ss - 2 m_q^2)^2 \,  M_{52} + 2 \epsilon m_q^2 (ss - 2 m_q^2) M_{50} \right. \nn\\
& & - \left.  2 \epsilon ss (ss - 2 m_q^2) M_{51}
+ 2 (ss - m_q^2) M_8 +4 (ss - m_q^2)^2 M_9  \right]  \, , \label{m510a4}\\
F_{53} & = & \, \epsilon^4 \, (ss - 2 m_q^2) \,  M_{53}  \, , \label{m512a1}\\
F_{54} & = & \, \epsilon^3 \, \sqrt{ ss}\sqrt{ss - 2 m_q^2}(ss - 2 m_q^2) \,  M_{54}  \, , \label{m512a2}\\
F_{55} & = & \, \epsilon^3 \, ss \left[ M_{55} - \epsilon M_{53} - (ss - 2 m_q^2)M_{54}/2 - M_{32}/2 \right]\, , \label{m512a3}\\
F_{56} & = & \, \epsilon^4 \, (ss - 2 m_q^2) \,  M_{56}  \, , \label{m517}\\
F_{57} & = & \, \epsilon^4 \, (ss - 2 m_q^2) \,  M_{57}  \, , \label{m518a1}\\
F_{58} & = & \, \epsilon^3 \, \sqrt{ ss}\sqrt{ss - 2 m_q^2}(ss - 2 m_q^2) \,  M_{58}  \, , \label{m518a2}\\
F_{59} & = & \, \epsilon^3 \,  ss \left[ M_{59} - \epsilon M_{57} - (ss - 2 m_q^2)M_{58}/2 - M_{42}/2 \right]  \, , \label{m518a3}\\
F_{60} & = & \, \epsilon^4 \, (ss - 2 m_q^2) \,  M_{60}  \, , \label{m523a1}
\eea
\bea
F_{61} & = & \, \epsilon^3 \, \sqrt{ ss}\sqrt{ss - 2 m_q^2}(ss - 2 m_q^2) \,  M_{61}  \, , \label{m523a2}\\
F_{62} & = & \, \epsilon^4 \, (ss - 2 m_q^2) \,  M_{62}  \, , \label{m531}\\
F_{63} & = & \, \epsilon^4 \, (ss - 2 m_q^2) \,  M_{63}  \, , \label{m532a1}\\
F_{64} & = & \, \epsilon^3 \, m_q^2 (ss - 2 m_q^2) \,  M_{64}  \, , \label{m532a2}\\
F_{65} & = & \, \epsilon^4 \, (ss - 2 m_q^2) \,  M_{65}  \, , \label{m533a1}\\
F_{66} & = & \, \epsilon^3 \, \sqrt{ ss}\sqrt{ss - 2 m_q^2}(ss - 2 m_q^2) \,  M_{66}  \, , \label{m533a2}\\
F_{67} & = & \, \epsilon^3 \, ss \left[ M_{67} - \epsilon M_{65} - (ss - 2 m_q^2)M_{66}/2 - M_{30}/2 \right]  \, , \label{m533a3}\\
F_{68} & = & \, \epsilon^4 \, (ss - 2 m_q^2) \,  M_{68}  \, , \label{m534}\\
F_{69} & = & \, \epsilon^4 \, (ss - 2 m_q^2) \,  M_{69}  \, , \label{m520}\\
F_{70} & = & \, \epsilon^4 \, (ss - 2 m_q^2) \,  M_{70}  \, , \label{m528}\\
F_{71} & = & \, \epsilon^4 \, (ss - 2 m_q^2)^2 \,  M_{71}  \, , \label{m68}\\
F_{72} & = & \, \epsilon^4 \, (ss - 2 m_q^2) \,  M_{72}  \, , \label{m614}\\
F_{73} & = & \, \epsilon^2 \,  ss (ss - 2 m_q^2) \,  M_{73}  \, , \label{m41}\\
F_{74} & = & \, \epsilon^2 \, (ss - m_q^2)^2 \,  M_{74}  \, , \label{m42}\\
F_{75} & = & \, \epsilon^2 \,  m_q^4 \,  M_{75}  \, , \label{m451}\\
F_{76} & = & \, \epsilon^3 \, (ss - 2 m_q^2) \,  M_{76}  \, , \label{m46}\\
F_{77} & = & \, \epsilon^3 \, (ss - 2 m_q^2) \,  M_{77}  \, , \label{m47}\\
F_{78} & = & \, \epsilon^3 \, (ss - 2 m_q^2) \,  M_{78}  \, , \label{m48}\\
F_{79} & = & \, \epsilon^3 \, (ss - m_q^2)(ss - 2 m_q^2) \,  M_{79}  \, , \label{m51}\\
F_{80} & = & \, \epsilon^3 \, (ss - m_q^2)(ss - 2 m_q^2) \,  M_{80}  \, , \label{m52}\\
F_{81} & = & \, \epsilon^3 \, m_q^2(ss - 2 m_q^2) \,  M_{81}  \, , \label{m53}\\
F_{82} & = & \, \epsilon^3 \, \sqrt{ ss}\sqrt{ss - 2 m_q^2}(ss - 2 m_q^2) \,  M_{82}  \, , \label{m54}\\
F_{83} & = & \, \epsilon^3 \, \sqrt{ ss}\sqrt{ss - 2 m_q^2}(ss - 2 m_q^2) \,  M_{83}  \, , \label{m55}\\
F_{84} & = & \, \epsilon^3 \, \sqrt{ ss}\sqrt{ss - 2 m_q^2}(ss - 2 m_q^2) \,  M_{84}  \, , \label{m56}\\
F_{85} & = & \, \epsilon^4 \, (ss - 2 m_q^2)^2 \,  M_{85}  \, , \label{m613}\\
F_{86} & = & \, \epsilon^3 \, 4 m_q^2(1-2\epsilon) \,  M_{86}  \, . \label{m526}
\eea

\vspace{3ex}

The integrals $M_1$ is defined as follow
\be
M_1 = \int {\mathcal D}^Dq_1 \, {\mathcal D}^Dq_2 \, \frac{1}{(-q_1^2+m_q^2)^2} \, \frac{1}{(-q_2^2+m_q^2)^2} = \frac{1}{\epsilon^2} \, ,
\ee
where the measure of the integration is defined as
\be
{\mathcal D}^Dq_i = \frac{1}{\pi^{D/2}\Gamma(1+\epsilon)}\left(\frac{m_q^2}{\mu^2}\right)^\epsilon  d^Dq_i \, .
\ee
For master integrals without numerators, their definition can be read off from Fig.\ref{midiag}, with the normalization defined above. For master integrals with numerators, we first define a series of propagators
\begin{align}
P_{1} & =m_{q}^{2}-q_{1}^{2},\hspace{3.1cm}P_{2}=m_{q}^{2}-q_{2}^{2},\nonumber \\
P_{3} & =-(q_{1}+q_{2})^{2},\hspace{2.52cm}P_{4}=m_{q}^{2}-(q_{1}+k_{1})^{2},\nonumber \\
P_{5} & =m_{q}^{2}-(q_{2}+k_{1})^{2},\hspace{1.82cm}P_{6}=m_{q}^{2}-(q_{1}+k_{2})^{2},\nonumber \\
P_{7} & =-(q_{1}+q_{2}+k_{1}+k_{2})^{2},\hspace{0.85cm}P_{8}=-(q_{1}+k_{q})^{2},\nonumber \\
P_{9} & =-(q_{2}+k_{q})^{2},\hspace{2.38cm}P_{10}=-(q_{1}-k_{q})^{2},\nonumber \\
P_{11} & =m_{q}^{2}-(q_{1}-2k_{q})^{2},\hspace{1.5cm}P_{12}=m_{q}^{2}-(q_{1}+2k_{q})^{2}.
\end{align}
Then, the master integrals with numerators can be expressed as
\bea
M_{26} & = & \int {\mathcal D}^Dq_1 \, {\mathcal D}^Dq_2 \frac{P_1}{P_2^2 P_3 P_6^2 P_8}, ~~~~~~~~
M_{40}  =  \int {\mathcal D}^Dq_1 \, {\mathcal D}^Dq_2 \frac{P_1}{P_2^2 P_3^2 P_6 P_{12}}, \nonumber\\
M_{55} & = & \int {\mathcal D}^Dq_1 \, {\mathcal D}^Dq_2 \frac{P_{9}}{P_1 P_2 P_3 P_5^2 P_{10}}, ~~~~
M_{59}  = \int {\mathcal D}^Dq_1 \, {\mathcal D}^Dq_2 \frac{P_{9}}{P_1 P_2 P_3 P_5^2 P_{11}}, \nonumber\\
M_{67} & = & \int {\mathcal D}^Dq_1 \, {\mathcal D}^Dq_2 \frac{P_{8}}{P_1 P_2 P_4^2 P_7 P_9},
\eea
and
\bea
M_{72}&=& \int {\mathcal D}^Dq_1 \, {\mathcal D}^Dq_2 \frac{(q_1+k_2)^2}{q_1^2 q_2^2 (q_{1}+q_{2})^{2} (q_1+q_2-k_2)^2(q_1+k_1)^2(2q_2\cdot k_q-q_2^2)}.
\eea

\section{Boundary conditions and analytic continuation}
Now, we are ready to perform the calculations of the differential equations. The first step is to specify all the boundary conditions that will completely fix the solution of the differential equations. Integrals $F_1, F_3,F_{11}, F_{14},F_{75}$ are constants that already known. They can easily be recalculated with the assistance of Mathematica packages {\bf MB} \cite{Czakon:2005rk} and {\bf AMBRE} \cite{Gluza:2007rt,Gluza:2010rn,Blumlein:2014maa}. The integrals $F_5, F_{10} , F_{15},F_{16}$ are regular at $ss=m_q^2$. Their boundary conditions at $ss=m_q^2$  can also be determined by employing packages {\bf MB}  and {\bf AMBRE}.

The integrals~$M_{(6\ldots9, 13, 19, 23, 51, 54, 55, 58, 59, 61,66, 67,73,82,83,84)}$~do not have singularity at~$ss=0$. Thanks  to their normalization factor~$ss$~ that multiplying with them in ${\textbf{F}}$, the corresponding canonical basis~$F_{(6\ldots9, 13, 19, 23, 51, 54, 55, 58, 59, 61,66, 67,73,82,83,84)}=0$, at~$ss=0$. Considering the fact that~$M_{(2, 4, 26, 29, 37, 74, 79, 80)}$~are regular at~$ss=m_q^2$, and their normalization factor to be $(ss - m_q^2)$, the boundary conditions of basis ~$F_{(2, 4, 26, 29, 37, 74, 79, 80)}$ can also be straightforwardly determined.

Since the integrals $M_{(18, 20\ldots28, 30\ldots36, 38, 39, 41\ldots50, 53, 56, 57, 60, 62\ldots65, 68\ldots72, 74, 76, 77, 78, 81, 85)}$ are regular at $ss = 2 m_q^2$, we can determinate boundary of the corresponding basis functions in ${\bf F}$ in a way similar to previous discussion.

For $M_{12}$ whose results are constants, we found that its results could be obtained by manipulating $F_{19}$. First the base $F_{19}$  equal to  $\epsilon^2 \, \left[ ss\, (ss - 2 m_q^2) \,  M_{19} + 3 ss/2 \, M_{12} \right]$, considering the fact that $M_{19}$ does not have singularity at $ss=2m_q^2$, and the normalization factor of $M_{19}$  in $F_{19}$ contains  $(ss - 2 m_q^2)$, we can take the limit at $ss\rightarrow 2 m_q^2$ for $F_{19}$ and obtain the results of $F_{12}$. With the assistance of PSLQ algorithm \cite{Ferguson:1999},  the results of $F_{12}$ can be simplified and expressed as follow
\be
F_{12}=\epsilon^2(\frac{\pi^2}{12} + \epsilon(\frac{7\zeta(3)}{4}-\frac{\pi^2\log(2)}{2})+ \epsilon^2(12\text{Li}_4(1/2)-\frac{31\pi^4}{360}+\pi^2\log^2(2)+\frac{\log^4(2)}{2})),
\ee
the results of weight four of $F_{12}$  is new to the best of our knowledge.

The result of $F_{17}$ can be obtained by taking the limit $ss\rightarrow 2 m_q^2$ for $F_{6}$ and expressed as
\be
F_{17}=\epsilon^2(\frac{\pi^2}{2} + \epsilon(\frac{21\zeta(3)}{2} + \pi^2\log(2))+ \epsilon^2(-24\text{Li}_4(1/2)+\frac{29\pi^4}{20}+2\pi^2\log^2(2)-\log^4(2))).
\ee
Because all the master integrals appeared in $F_{40}$ are regular at $ss = \frac{8 m_q^2}{3}$, the boundary of $F_{40}$ is calculated by taking the limit $ss\rightarrow \frac{8 m_q^2}{3}$ for $F_{40}$.  Considering the fact that $M_{49}$ does not have singularity at $ss = 0$, the boundary condition of $F_{52}$ can be determined from the differential equation of $F_{49}$. To further illustrate it, we consider the differential of $F_{49}$ with respect to variable $x$ and find that
\be
\frac{\partial F_{49}}{\partial x}= \epsilon(\frac{1}{x-1})\left(F_4-F_3+2 F_6 + F_{27} - 2 F_{49} + F_{50}+ F_{52} \right)+\ldots ,\label{de49}
\ee
where ellipses stand for less singular terms. All the integrals that appear in Eq.(\ref{de49}) have finite limits at $x \rightarrow 1(ss \rightarrow 0)$. This consistency  leads to a relation between different integrals
\be
\lim_{x \rightarrow 1} F_4-F_3+2 F_6 + F_{27} - 2 F_{49} + F_{50}+ F_{52} =0.
\ee
We can then obtain the boundary condition of $F_{52}$ at $x=1$ from the above equation.

A more general calculation of the kinematics of $F_{86}$ was performed before \cite{Broadhurst,Davydychev:2003mv,Bonciani:2016qxi}. By taking the limit $p^2=4 m^2$ for $F_{10101}$ in \cite{Davydychev:2003mv}, and by adopting the PSLQ algorithm, we obtain the results of $F_{86}$
\bea
F_{86}&=& \epsilon^3(-\frac{21\zeta(3)}{2}+2\pi^2\log(2)+\frac{i \pi^3 }{2}+\epsilon \left[ -7920\text{Li}_4(1/2) - 1260\zeta(3)\log(2)\right. \nn\\
& & + \left. 330\pi^2\log^2(2)-330\log^4(2)-4\pi^4 + i \pi(630\zeta(3)-90\pi^2\log(2))  \right]/90).
\eea
By now, all the boundary conditions are fixed.

The next step is to determinate the analytic continuations of the master integrals. When we consider the processes (\ref{pro1},\ref{pro2}), the variables of the master integrals lie either in Euclidean region or in Minkowski region. Their analytic continuations should be considered carefully. The proper analytic continuation can be achieved by the replacement $ss \rightarrow ss + i 0$ at fixed $m_q^2$. This transfer corresponds to $x\rightarrow x + i 0, y\rightarrow y + i 0$ and $z\rightarrow z + i 0$. For single scale integrals that depend only on $m_q^2$, the replacement $m_q^2 \rightarrow m_q^2 + i 0$ is sufficient for obtaining the correct results.

The calculations are performed with our self-written Mathematica code. All the analytical expressions of the master integrals require an independent examination. We check all the results against the results obtained from numerical programs Fiesta \cite{Smirnov:2013eza,Smirnov:2015mct} and SecDec \cite{Borowka:2012yc,Borowka:2015mxa}. Good agreement has been achieved between the analytical and numerical approaches with kinematics in both Euclidean region and Minkowski region.

\section{Discussions and Conclusions}
In this work, we obtained the analytic results of 86 out of the 133 master integrals involved in the calculation of NNLO corrections to CP-even heavy quarkonium production processes such as~$\gamma^*\gamma\rightarrow Q\bar{Q}$~and~$e^+e^-\rightarrow \gamma+ Q\bar{Q}$. By choosing a proper canonical basis, the differential equation group is cast into a canonical form. All of the 86 master integrals are then expressed in terms of Harmonic polylogarithms and Goncharov polylogarithms. The integrals obtained here may also be applied to the calculation of NNLO corrections of other process, such as the exclusive decay of higgs boson or $Z_0$ boson into CP-even quarkonium plus a photon and the inclusive hadron production or decay of $\eta_c/\eta_b$ . The remaining integrals which cannot be expressed in Goncharov polylogarithms require a further investigation.

\acknowledgments

This work was supported in part by the Ministry of Science and Technology of the People's Republic of China(2015CB856703); by the Strategic Priority Research Program of the Chinese Academy of Sciences, Grant No.XDB23030100; and by the National Natural Science Foundation of China(NSFC) under the grants 11375200 and 11635009.

\appendix


\section{The analytical results}
Here we list the analytical results of all the 86 canonical basis $F_i$ up to weight 3. They are expressed in terms of HPLs and GPLs.

\allowdisplaybreaks{
\begin{flalign}
 & F_{1}=1,\\
 & F_{2}=\epsilon[-H_{0}(y)+i\pi]+\epsilon^{2}[-H_{-1,0}(y)+2H_{0,0}(y)+i\pi H_{-1}(y)-2i\pi H_{0}(y)-\frac{5\pi^{2}}{6}]\nonumber \\
 & \epsilon^{3}[i\pi H_{-1,-1}(y)-2i\pi H_{-1,0}(y)-2i\pi H_{0,-1}(y)+4i\pi H_{0,0}(y)-H_{-1,-1,0}(y)\nonumber \\
 & +2H_{-1,0,0}(y)+2H_{0,-1,0}(y)-4H_{0,0,0}(y)-\frac{5}{6}\pi^{2}H_{-1}(y)+\frac{5}{3}\pi^{2}H_{0}(y)-\zeta(3)-\frac{i\pi^{3}}{3}]+{\cal O}(\epsilon^{4}),\\
 & F_{3}=1/4+\epsilon^{2}[\frac{\pi^{2}}{6}]+\epsilon^{3}[2\zeta(3)]+{\cal O}(\epsilon^{4}),\\
 & F_{4}=\epsilon[H_{0}(y)-i\pi]+\epsilon^{2}[H_{-1,0}(y)-4H_{0,0}(y)-i\pi H_{-1}(y)+4i\pi H_{0}(y)+\frac{11\pi^{2}}{6}]\nonumber \\
 & +\epsilon^{3}[-i\pi H_{-1,-1}(y)+4i\pi H_{-1,0}(y)+6i\pi H_{0,-1}(y)-16i\pi H_{0,0}(y)+H_{-1,-1,0}(y)\nonumber \\
 & -4H_{-1,0,0}(y)-6H_{0,-1,0}(y)+16H_{0,0,0}(y)\nonumber \\
 & +\frac{11}{6}\pi^{2}H_{-1}(y)-\frac{20}{3}\pi^{2}H_{0}(y)+7\zeta(3)+\frac{4i\pi^{3}}{3}]+{\cal O}(\epsilon^{4}),\\
 & F_{5}=-1+\epsilon[2\left(H_{0}(y)-i\pi\right)]\nonumber \\
 & +\epsilon^{2}[4H_{-1,0}(y)-8H_{0,0}(y)-4i\pi H_{-1}(y)+8i\pi H_{0}(y)+3\pi^{2}]\nonumber \\
 & +\epsilon^{3}\frac{2}{3}[-6i\pi H_{-1,-1}(y)+24i\pi H_{-1,0}(y)+18i\pi H_{0,-1}(y)-48i\pi H_{0,0}(y)+6H_{-1,-1,0}(y)\nonumber \\
 & -24H_{-1,0,0}(y)-18H_{0,-1,0}(y)+48H_{0,0,0}(y)\nonumber \\
 & +11\pi^{2}H_{-1}(y)-20\pi^{2}H_{0}(y)+9\zeta(3)+4i\pi^{3}]+{\cal O}(\epsilon^{4}),\\
 & F_{6}=\epsilon^{2}[-G_{0,0}(x)]+\epsilon^{3}[6G_{0,-1,0}(x)-3G_{0,0,0}(x)+2G_{0,1,0}(x)-2G_{1,0,0}(x)\nonumber \\
 & +\frac{1}{6}\pi^{2}G_{0}(x)+3\zeta(3)]+{\cal O}(\epsilon^{4}),\\
 & F_{7}=\epsilon\left(-\frac{1}{2}G_{0}(x)\right)+\epsilon^{2}\left(3G_{-1,0}(x)-2G_{0,0}(x)+G_{1,0}(x)+\frac{\pi^{2}}{12}\right)\nonumber \\
 & +\epsilon^{3}[-18G_{-1,-1,0}(x)+12G_{-1,0,0}(x)-6G_{-1,1,0}(x)+12G_{0,-1,0}(x)-5G_{0,0,0}(x)\nonumber \\
 & +4G_{0,1,0}(x)-6G_{1,-1,0}(x)+4G_{1,0,0}(x)-2G_{1,1,0}(x)-\frac{1}{2}\pi^{2}G_{-1}(x)\nonumber \\
 & +\frac{1}{3}\pi^{2}G_{0}(x)-\frac{1}{6}\pi^{2}G_{1}(x)+\frac{11\zeta(3)}{2}]+{\cal O}(\epsilon^{4}),\\
 & F_{8}=\epsilon[\frac{H_{0}(x)}{2}]\nonumber \\
 & +\epsilon^{2}[\frac{1}{12}\left(6H_{0,0}(x)-\pi^{2}\right)-H_{-1,0}(x)]\nonumber \\
 & +\epsilon^{3}[2H_{-1,-1,0}(x)-H_{-1,0,0}(x)-H_{0,-1,0}(x)\nonumber \\
 & +\frac{1}{2}H_{0,0,0}(x)+\frac{1}{6}\pi^{2}H_{-1}(x)-\frac{1}{12}\pi^{2}H_{0}(x)-\zeta(3)]+{\cal O}(\epsilon^{4}),\\
 & F_{9}=\epsilon^{2}[\frac{1}{2}H_{0}(x)\left(-H_{0}(y)+i\pi\right)]+\epsilon^{3}[\left(-H_{0}(y)+i\pi\right)\left(-H_{-1,0}(x)+\frac{1}{2}H_{0,0}(x)-\frac{\pi^{2}}{12}\right)\nonumber \\
 & +\frac{1}{2}H_{0}(x)\left(-H_{-1,0}(y)+2H_{0,0}(y)+i\pi H_{-1}(y)-2i\pi H_{0}(y)-\frac{5\pi^{2}}{6}\right)]+{\cal O}(\epsilon^{4}),\\
 & F_{10}=\frac{1}{2}+\epsilon[\frac{1}{2}i\left(iH_{-1}(z)+\pi+i\log(2)\right)]\nonumber \\
 & +\epsilon^{2}[\frac{1}{12}\left(6H_{-1,-1}(z)+6(\log(2)-i\pi)H_{-1}(z)-4\pi^{2}+3\log^{2}(2)-6i\pi\log(2)\right)]\nonumber \\
 & +\epsilon^{3}[\frac{1}{12}\big(-6H_{-1,-1,-1}(z)+6i(\pi+i\log(2))H_{-1,-1}(z)\nonumber \\
 & +\left(4\pi^{2}-3\log^{2}(2)+6i\pi\log(2)\right)H_{-1}(z)\nonumber \\
 & -12\zeta(3)-2i\pi^{3}-\log^{3}(2)+3i\pi\log^{2}(2)+\pi^{2}\log(16)\big)]+{\cal O}(\epsilon^{4}),\\
 & F_{11}=\frac{1}{2}+\epsilon[-\log(2)+\frac{i\pi}{2}]+\epsilon^{2}[-\frac{\pi^{2}}{3}+\log^{2}(2)-i\pi\log(2)]\nonumber \\
 & +\epsilon^{3}[\frac{1}{6}\left(-6\zeta(3)-i\pi^{3}-4\log^{3}(2)+6i\pi\log^{2}(2)+4\pi^{2}\log(2)\right)]+{\cal O}(\epsilon^{4}),\\
 & F_{12}=\epsilon^{2}[\frac{\pi^{2}}{12}]+\epsilon^{3}[\frac{7\zeta(3)}{4}-\frac{1}{2}\pi^{2}\log(2)]+{\cal O}(\epsilon^{4}),\\
 & F_{13}=\epsilon[\frac{H_{0}(x)}{8}]\nonumber \\
 & +\epsilon^{2}[\frac{1}{48}\left(-12H_{-1,0}(x)+6H_{0,0}(x)+6i\pi H_{0}(x)-12\log(2)H_{0}(x)-\pi^{2}\right)]\nonumber \\
 & +\epsilon^{3}[\frac{1}{48}(24H_{-1,-1,0}(x)-12H_{-1,0,0}(x)-12H_{0,-1,0}(x)\nonumber \\
 & +6H_{0,0,0}(x)+2\pi^{2}H_{-1}(x)-\pi^{2}H_{0}(x)-12\zeta(3))\nonumber \\
 & +\left(-\frac{\log(2)}{2}+\frac{i\pi}{4}\right)\left(-H_{-1,0}(x)+\frac{1}{2}H_{0,0}(x)-\frac{\pi^{2}}{12}\right)\nonumber \\
 & +\frac{1}{12}\left(-\pi^{2}+3\log^{2}(2)-3i\pi\log(2)\right)H_{0}(x)]+{\cal O}(\epsilon^{4}),\\
 & F_{14}=-\frac{1}{4}+\epsilon[\log(2)-\frac{i\pi}{2}]+\epsilon^{2}[\frac{7\pi^{2}}{12}-2\log^{2}(2)+2i\pi\log(2)]\nonumber \\
 & +\epsilon^{3}[\frac{1}{6}\left(15\zeta(3)+3i\pi^{3}+16\log^{3}(2)-24i\pi\log^{2}(2)-14\pi^{2}\log(2)\right)]+{\cal O}(\epsilon^{4}),\\
 & F_{15}=-\frac{1}{2}+\epsilon\left(H_{-1}(z)-i\pi+\log(2)\right)\nonumber \\
 & +\epsilon^{2}\left(-2H_{-1,-1}(z)+2i(\pi+i\log(2))H_{-1}(z)+\frac{7\pi^{2}}{6}-\log^{2}(2)+2i\pi\log(2)\right)\nonumber \\
 & +\epsilon^{3}[\frac{1}{3}\big(12H_{-1,-1,-1}(z)+12(\log(2)-i\pi)H_{-1,-1}(z)\nonumber \\
 & +\left(-7\pi^{2}+6\log^{2}(2)-12i\pi\log(2)\right)H_{-1}(z)\nonumber \\
 & +15\zeta(3)+3i\pi^{3}+2\log^{3}(2)-6i\pi\log^{2}(2)-7\pi^{2}\log(2)\big)]+{\cal O}(\epsilon^{4}),\\
 & F_{16}=\frac{1}{8}+\epsilon[\frac{1}{8}i\left(iH_{-1}(z)+2\pi+3i\log(2)\right)]\nonumber \\
 & +\epsilon^{2}[\frac{1}{48}\left(6H_{-1,-1}(z)+6(\log(8)-2i\pi)H_{-1}(z)-14\pi^{2}+3\log^{2}(8)-6i\pi\log(64)\right)]\nonumber \\
 & +\epsilon^{3}[\frac{1}{48}\big(-6H_{-1,-1,-1}(z)+(-6\log(8)+12i\pi)H_{-1,-1}(z)\nonumber \\
 & +\left(14\pi^{2}-3\log^{2}(8)+6i\pi\log(64)\right)H_{-1}(z)-24\zeta(3)\nonumber \\
 & -12i\pi^{3}-\log^{3}(8)+54i\pi\log^{2}(2)+14\pi^{2}\log(8)\big)]+{\cal O}(\epsilon^{4}),\\
 & F_{17}=\epsilon^{2}[\frac{\pi^{2}}{2}]+\epsilon^{3}[\frac{21\zeta(3)}{2}+\pi^{2}\log(2)]+{\cal O}(\epsilon^{4}),\\
 & F_{18}=\epsilon^{3}[H_{0,0,1}(z)-\zeta(3)]+{\cal O}(\epsilon^{4}),\\
 & F_{19}=\epsilon^{2}[H_{0,1}(z)+\frac{\pi^{2}}{12}]\nonumber \\
 & +\epsilon^{3}[-4H_{-1,0,1}(z)+4H_{0,0,1}(z)+2H_{0,1,1}(z)+2H_{1,0,1}(z)\nonumber \\
 & -\frac{1}{3}\pi^{2}H_{-1}(z)-\frac{1}{3}\pi^{2}H_{1}(z)+\frac{3\zeta(3)}{4}-\frac{1}{2}\pi^{2}\log(2)]+{\cal O}(\epsilon^{4}),\\
 & F_{20}=\epsilon^{3}[2G_{0,0,-1}(x)-2G_{0,-i,-1}(x)+G_{0,-i,0}(x)-2G_{0,i,-1}(x)+G_{0,i,0}(x)\nonumber \\
 & -2G_{-i,0,-1}(x)-G_{-i,0,0}(x)+2G_{-i,-i,-1}(x)-G_{-i,-i,0}(x)+2G_{-i,i,-1}(x)-G_{-i,i,0}(x)\nonumber \\
 & -2G_{i,0,-1}(x)-G_{i,0,0}(x)+2G_{i,-i,-1}(x)-G_{i,-i,0}(x)+2G_{i,i,-1}(x)-G_{i,i,0}(x)+2G_{1,0,0}(x)\nonumber \\
 & -\log(2)G_{0,0}(x)+\log(2)G_{0,-i}(x)+\log(2)G_{0,i}(x)+\log(2)G_{-i,0}(x)-\log(2)G_{-i,-i}(x)\nonumber \\
 & -\log(2)G_{-i,i}(x)+\log(2)G_{i,0}(x)-\log(2)G_{i,-i}(x)-\log(2)G_{i,i}(x)+\frac{1}{12}\pi^{2}G_{0}(x)\nonumber \\
 & -\frac{1}{12}\pi^{2}G_{-i}(x)-\frac{1}{12}\pi^{2}G_{i}(x)-\frac{1}{2}\log^{2}(2)G_{0}(x)+\frac{1}{2}\log^{2}(2)G_{-i}(x)+\frac{1}{2}\log^{2}(2)G_{i}(x)\nonumber \\
 & +\frac{1}{24}\left(-3\zeta(3)-4\log^{3}(2)+\pi^{2}\log(4)\right)]+{\cal O}(\epsilon^{4}),\\
 & F_{21}=\epsilon^{2}[-2G_{0,-1}(x)+2G_{-i,-1}(x)-G_{-i,0}(x)+2G_{i,-1}(x)-G_{i,0}(x)\nonumber \\
 & +\log(2)G_{0}(x)-\log(2)G_{-i}(x)-\log(2)G_{i}(x)+\frac{1}{6}\left(3\log^{2}(2)-\pi^{2}\right)]\nonumber \\
 & +\epsilon^{3}[16G_{-1,0,-1}(x)-16G_{-1,-i,-1}(x)+8G_{-1,-i,0}(x)-16G_{-1,i,-1}(x)+8G_{-1,i,0}(x)\nonumber \\
 & +8G_{0,-1,-1}(x)+2G_{0,-1,0}(x)-8G_{0,0,-1}(x)+4G_{0,-i,-1}(x)-2G_{0,-i,0}(x)+4G_{0,i,-1}(x)\nonumber \\
 & -2G_{0,i,0}(x)+2G_{0,1,0}(x)-8G_{-i,-1,-1}(x)+4G_{-i,-1,0}(x)-4G_{-i,0,-1}(x)-4G_{-i,0,0}(x)\nonumber \\
 & +8G_{-i,-i,-1}(x)-4G_{-i,-i,0}(x)+8G_{-i,i,-1}(x)-4G_{-i,i,0}(x)-8G_{i,-1,-1}(x)+4G_{i,-1,0}(x)\nonumber \\
 & -4G_{i,0,-1}(x)-4G_{i,0,0}(x)+8G_{i,-i,-1}(x)-4G_{i,-i,0}(x)+8G_{i,i,-1}(x)-4G_{i,i,0}(x)\nonumber \\
 & +8G_{1,0,-1}(x)-8G_{1,-i,-1}(x)+4G_{1,-i,0}(x)-8G_{1,i,-1}(x)+4G_{1,i,0}(x)\nonumber \\
 & -8\log(2)G_{-1,0}(x)+8\log(2)G_{-1,-i}(x)+8\log(2)G_{-1,i}(x)-4\log(2)G_{0,-1}(x)\nonumber \\
 & +4\log(2)G_{0,0}(x)-2\log(2)G_{0,-i}(x)-2\log(2)G_{0,i}(x)+4\log(2)G_{-i,-1}(x)\nonumber \\
 & +2\log(2)G_{-i,0}(x)-4\log(2)G_{-i,-i}(x)-4\log(2)G_{-i,i}(x)+4\log(2)G_{i,-1}(x)\nonumber \\
 & +2\log(2)G_{i,0}(x)-4\log(2)G_{i,-i}(x)-4\log(2)G_{i,i}(x)-4\log(2)G_{1,0}(x)+4\log(2)G_{1,-i}(x)\nonumber \\
 & +4\log(2)G_{1,i}(x)+\frac{4}{3}\pi^{2}G_{-1}(x)-\frac{1}{4}\pi^{2}G_{0}(x)-\frac{1}{12}\pi^{2}G_{-i}(x)-\frac{1}{12}\pi^{2}G_{i}(x)-\frac{1}{3}\pi^{2}G_{1}(x)\nonumber \\
 & -4\log^{2}(2)G_{-1}(x)+2\log^{2}(2)G_{0}(x)+\log^{2}(2)G_{-i}(x)+\log^{2}(2)G_{i}(x)-2\log^{2}(2)G_{1}(x)\nonumber \\
 & +\frac{1}{24}\left(-15\zeta(3)+16\log^{3}(2)+\pi^{2}\log(4)\right)]+{\cal O}(\epsilon^{4}),\\
 & F_{22}=\epsilon^{2}[2G_{0,-1}(x)-2G_{-i,-1}(x)+G_{-i,0}(x)-2G_{i,-1}(x)+G_{i,0}(x)\nonumber \\
 & -\log(2)G_{0}(x)+\log(2)G_{-i}(x)+\log(2)G_{i}(x)+\frac{1}{6}\left(\pi^{2}-3\log^{2}(2)\right)]\nonumber \\
 & +\epsilon^{3}[-4G_{-1,0,-1}(x)+4G_{-1,-i,-1}(x)-2G_{-1,-i,0}(x)+4G_{-1,i,-1}(x)-2G_{-1,i,0}(x)\nonumber \\
 & -16G_{0,-1,-1}(x)+4G_{0,-1,0}(x)+4G_{0,0,-1}(x)+2G_{0,-i,-1}(x)-G_{0,-i,0}(x)+2G_{0,i,-1}(x)\nonumber \\
 & -G_{0,i,0}(x)+G_{0,1,0}(x)+16G_{-i,-1,-1}(x)-8G_{-i,-1,0}(x)-6G_{-i,0,-1}(x)+3G_{-i,0,0}(x)\nonumber \\
 & -2G_{-i,-i,-1}(x)+G_{-i,-i,0}(x)-2G_{-i,i,-1}(x)+G_{-i,i,0}(x)+16G_{i,-1,-1}(x)-8G_{i,-1,0}(x)\nonumber \\
 & -6G_{i,0,-1}(x)+3G_{i,0,0}(x)-2G_{i,-i,-1}(x)+G_{i,-i,0}(x)-2G_{i,i,-1}(x)+G_{i,i,0}(x)\nonumber \\
 & +4G_{1,0,-1}(x)-G_{1,0,0}(x)-4G_{1,-i,-1}(x)+2G_{1,-i,0}(x)-4G_{1,i,-1}(x)+2G_{1,i,0}(x)\nonumber \\
 & +2\log(2)G_{-1,0}(x)-2\log(2)G_{-1,-i}(x)-2\log(2)G_{-1,i}(x)+8\log(2)G_{0,-1}(x)\nonumber \\
 & -2\log(2)G_{0,0}(x)-\log(2)G_{0,-i}(x)-\log(2)G_{0,i}(x)-8\log(2)G_{-i,-1}(x)+3\log(2)G_{-i,0}(x)\nonumber \\
 & +\log(2)G_{-i,-i}(x)+\log(2)G_{-i,i}(x)-8\log(2)G_{i,-1}(x)+3\log(2)G_{i,0}(x)+\log(2)G_{i,-i}(x)\nonumber \\
 & +\log(2)G_{i,i}(x)-2\log(2)G_{1,0}(x)+2\log(2)G_{1,-i}(x)+2\log(2)G_{1,i}(x)-\frac{1}{6}\pi^{2}G_{-i}(x)\nonumber \\
 & -\frac{1}{6}\pi^{2}G_{i}(x)-\frac{1}{6}\pi^{2}G_{1}(x)+\left(\log^{2}(2)-\frac{\pi^{2}}{3}\right)G_{-1}(x)+\frac{1}{12}\left(\pi^{2}-18\log^{2}(2)\right)G_{0}(x)\nonumber \\
 & +\frac{3}{2}\log^{2}(2)G_{-i}(x)+\frac{3}{2}\log^{2}(2)G_{i}(x)-\log^{2}(2)G_{1}(x)\nonumber \\
 & +\frac{23\zeta(3)}{8}-\frac{\log^{3}(2)}{2}+\frac{2}{3}\pi^{2}\log(2)]+{\cal O}(\epsilon^{4}),\\
 & F_{23}=\epsilon^{2}[-4G_{-1,0}(x)-2G_{0,-1}(x)+\frac{5}{2}G_{0,0}(x)+G_{1,0}(x)+\log(2)G_{0}(x)-\frac{\pi^{2}}{3}]\nonumber \\
 & +\epsilon^{3}[16G_{-1,-1,0}(x)+8G_{-1,0,-1}(x)-10G_{-1,0,0}(x)-4G_{-1,1,0}(x)+16G_{0,-1,-1}(x)\nonumber \\
 & -14G_{0,-1,0}(x)-5G_{0,0,-1}(x)+\frac{11}{2}G_{0,0,0}(x)-6G_{0,-i,-1}(x)+3G_{0,-i,0}(x)\nonumber \\
 &-6G_{0,i,-1}(x) +3G_{0,i,0}(x)+\frac{3}{2}G_{0,1,0}(x)-4G_{1,-1,0}(x)-2G_{1,0,-1}(x)+\frac{5}{2}G_{1,0,0}(x)\nonumber \\
 & +G_{1,1,0}(x)-4\log(2)G_{-1,0}(x)-8\log(2)G_{0,-1}(x)+\frac{5}{2}\log(2)G_{0,0}(x)+3\log(2)G_{0,-i}(x)\nonumber \\
 & +3\log(2)G_{0,i}(x)+\log(2)G_{1,0}(x)+\frac{4}{3}\pi^{2}G_{-1}(x)-\frac{1}{3}\pi^{2}G_{1}(x)+\left(\frac{\log^{2}(2)}{2}-\frac{7\pi^{2}}{24}\right)G_{0}(x)\nonumber \\
 & -\frac{169\zeta(3)}{16}-\frac{1}{3}\pi^{2}\log(2)]+{\cal O}(\epsilon^{4}),\\
 & F_{24}=\epsilon^{3}[-i\pi H_{-1,-1}(y)+i\pi H_{-1,0}(y)-i\pi H_{1,0}(y)+H_{-1,-1,0}(y)-H_{-1,0,0}(y)+H_{1,0,0}(y)\nonumber \\
 & +\frac{1}{3}\pi^{2}H_{-1}(y)]+{\cal O}(\epsilon^{4}),\\
 & F_{25}=\epsilon^{2}[\frac{1}{2}i\left(\pi H_{-1}(y)+iH_{-1,0}(y)\right)]\nonumber \\
 & +\epsilon^{3}[\frac{1}{2}i\pi H_{-1,-1}(y)-i\pi H_{-1,0}(y)-i\pi H_{0,-1}(y)-i\pi H_{1,0}(y)-\frac{1}{2}H_{-1,-1,0}(y)\nonumber \\
 & +H_{-1,0,0}(y)+H_{0,-1,0}(y)+H_{1,0,0}(y)-\frac{5}{12}\pi^{2}H_{-1}(y)]+{\cal O}(\epsilon^{4}),\\
 & F_{26}=\epsilon^{2}[\frac{1}{3}\left(6H_{-1,0}(y)-6H_{0,0}(y)-6i\pi H_{-1}(y)+6i\pi H_{0}(y)+\pi^{2}\right)]\nonumber \\
 & +\epsilon^{3}[-6i\pi H_{-1,-1}(y)+8i\pi H_{-1,0}(y)+8i\pi H_{0,-1}(y)-12i\pi H_{0,0}(y)+6H_{-1,-1,0}(y)\nonumber \\
 &-8H_{-1,0,0}(y) -8H_{0,-1,0}(y) +12H_{0,0,0}(y)+3\pi^{2}H_{-1}(y)-\frac{11}{3}\pi^{2}H_{0}(y)-2\zeta(3)-\frac{i\pi^{3}}{3}]\nonumber \\
 &+{\cal O}(\epsilon^{4}),\\
 & F_{27}=\epsilon^{3}[-\frac{1}{2}\log^{2}(2)G_{0}(x)+\frac{1}{12}\pi^{2}G_{0}(x)+\frac{1}{2}\log^{2}(2)G_{-i}(x)+\frac{1}{3}\pi^{2}G_{-i}(x)+\frac{1}{2}\log^{2}(2)G_{i}(x)\nonumber \\ & +\frac{1}{3}\pi^{2}G_{i}(x)-\frac{5}{12}\pi^{2}G_{\frac{1}{2}\left(-1-i\sqrt{3}\right)}(x)-\frac{5}{12}\pi^{2}G_{\frac{1}{2}\left(-1+i\sqrt{3}\right)}(x)+2\log(2)G_{0,-1}(x)\nonumber \\
 & -\log(2)G_{0,0}(x)-4\log(2)G_{-i,-1}(x)+\log(2)G_{-i,0}(x)+\log(2)G_{-i,-i}(x)\nonumber \\
 &+\log(2)G_{-i,i}(x) -4\log(2)G_{i,-1}(x)+\log(2)G_{i,0}(x)+\log(2)G_{i,-i}(x)+\log(2)G_{i,i}(x)\nonumber \\
 & +2\log(2)G_{\frac{1}{2}\left(-1-i\sqrt{3}\right),-1}(x)-\log(2)G_{\frac{1}{2}\left(-1-i\sqrt{3}\right),-i}(x)-\log(2)G_{\frac{1}{2}\left(-1-i\sqrt{3}\right),i}(x)\nonumber \\
 & +2\log(2)G_{\frac{1}{2}\left(-1+i\sqrt{3}\right),-1}(x)-\log(2)G_{\frac{1}{2}\left(-1+i\sqrt{3}\right),-i}(x)-\log(2)G_{\frac{1}{2}\left(-1+i\sqrt{3}\right),i}(x)\nonumber \\
 & -4G_{0,-1,-1}(x)+2G_{0,-1,0}(x)+2G_{0,0,-1}(x)+8G_{-i,-1,-1}(x)-4G_{-i,-1,0}(x)\nonumber \\
 & -2G_{-i,0,-1}(x)+5G_{-i,0,0}(x)-2G_{-i,-i,-1}(x)+G_{-i,-i,0}(x)-2G_{-i,i,-1}(x)+G_{-i,i,0}(x)\nonumber \\
 & +8G_{i,-1,-1}(x)-4G_{i,-1,0}(x)-2G_{i,0,-1}(x)+5G_{i,0,0}(x)-2G_{i,-i,-1}(x)+G_{i,-i,0}(x)\nonumber \\
 & -2G_{i,i,-1}(x)+G_{i,i,0}(x)-4G_{1,0,0}(x)-4G_{\frac{1}{2}\left(-1-i\sqrt{3}\right),-1,-1}(x)\nonumber \\
 & +2G_{\frac{1}{2}\left(-1-i\sqrt{3}\right),-1,0}(x)-3G_{\frac{1}{2}\left(-1-i\sqrt{3}\right),0,0}(x)+2G_{\frac{1}{2}\left(-1-i\sqrt{3}\right),-i,-1}(x)-G_{\frac{1}{2}\left(-1-i\sqrt{3}\right),-i,0}(x)\nonumber \\
 & +2G_{\frac{1}{2}\left(-1-i\sqrt{3}\right),i,-1}(x)-G_{\frac{1}{2}\left(-1-i\sqrt{3}\right),i,0}(x)-4G_{\frac{1}{2}\left(-1+i\sqrt{3}\right),-1,-1}(x)+2G_{\frac{1}{2}\left(-1+i\sqrt{3}\right),-1,0}(x)\nonumber \\
 & -3G_{\frac{1}{2}\left(-1+i\sqrt{3}\right),0,0}(x)+2G_{\frac{1}{2}\left(-1+i\sqrt{3}\right),-i,-1}(x)-G_{\frac{1}{2}\left(-1+i\sqrt{3}\right),-i,0}(x)+2G_{\frac{1}{2}\left(-1+i\sqrt{3}\right),i,-1}(x)\nonumber \\
 & -G_{\frac{1}{2}\left(-1+i\sqrt{3}\right),i,0}(x)+\frac{1}{24}\left(-8\pi^{2}\log(2)-4\log^{3}(2)+3\zeta(3)\right)]+{\cal O}(\epsilon^{4}),\\
 & F_{28}=\epsilon^{2}[-G_{0,-1}(x)+G_{-i,-1}(x)-\frac{1}{2}G_{-i,0}(x)+G_{i,-1}(x)-\frac{1}{2}G_{i,0}(x)\nonumber \\
 & +\frac{1}{2}\log(2)G_{0}(x)-\frac{1}{2}\log(2)G_{-i}(x)-\frac{1}{2}\log(2)G_{i}(x)+\frac{1}{12}\left(3\log^{2}(2)-\pi^{2}\right)]\nonumber \\
 & +\epsilon^{3}[-\log^{2}(2)G_{-1}(x)+\frac{1}{3}\pi^{2}G_{-1}(x)+\frac{1}{4}\log^{2}(2)G_{0}(x)+\frac{1}{4}\log^{2}(2)G_{-i}(x)+\frac{1}{3}\pi^{2}G_{-i}(x)\nonumber \\
 & +\frac{1}{4}\log^{2}(2)G_{i}(x)+\frac{1}{3}\pi^{2}G_{i}(x)-\frac{5}{12}\pi^{2}G_{\frac{1}{2}\left(-1-i\sqrt{3}\right)}(x)-\frac{5}{12}\pi^{2}G_{\frac{1}{2}\left(-1+i\sqrt{3}\right)}(x)\nonumber \\
 & -2\log(2)G_{-1,0}(x)+2\log(2)G_{-1,-i}(x)+2\log(2)G_{-1,i}(x)-2\log(2)G_{0,-1}(x)+\nonumber \\
 & \frac{1}{2}\log(2)G_{0,0}(x)+\frac{1}{2}\log(2)G_{0,-i}(x)+\frac{1}{2}\log(2)G_{0,i}(x)+\frac{1}{2}\log(2)G_{-i,0}(x)\nonumber \\
 & -\frac{1}{2}\log(2)G_{-i,-i}(x)-\frac{1}{2}\log(2)G_{-i,i}(x)+\frac{1}{2}\log(2)G_{i,0}(x)-\frac{1}{2}\log(2)G_{i,-i}(x)\nonumber \\
 & -\frac{1}{2}\log(2)G_{i,i}(x)+2\log(2)G_{\frac{1}{2}\left(-1-i\sqrt{3}\right),-1}(x)-\log(2)G_{\frac{1}{2}\left(-1-i\sqrt{3}\right),-i}(x)\nonumber \\
 & -\log(2)G_{\frac{1}{2}\left(-1-i\sqrt{3}\right),i}(x)+2\log(2)G_{\frac{1}{2}\left(-1+i\sqrt{3}\right),-1}(x)-\log(2)G_{\frac{1}{2}\left(-1+i\sqrt{3}\right),-i}(x)\nonumber \\
 & -\log(2)G_{\frac{1}{2}\left(-1+i\sqrt{3}\right),i}(x)+4G_{-1,0,-1}(x)-4G_{-1,-i,-1}(x)+2G_{-1,-i,0}(x)-4G_{-1,i,-1}(x)\nonumber \\
 & +2G_{-1,i,0}(x)+4G_{0,-1,-1}(x)+G_{0,-1,0}(x)-G_{0,0,-1}(x)-G_{0,-i,-1}(x)+\frac{1}{2}G_{0,-i,0}(x)\nonumber \\
 & -G_{0,i,-1}(x)+\frac{1}{2}G_{0,i,0}(x)+G_{0,1,0}(x)-G_{-i,0,-1}(x)+\frac{5}{2}G_{-i,0,0}(x)+G_{-i,-i,-1}(x)\nonumber \\
 & -\frac{1}{2}G_{-i,-i,0}(x)+G_{-i,i,-1}(x)-\frac{1}{2}G_{-i,i,0}(x)-G_{i,0,-1}(x)+\frac{5}{2}G_{i,0,0}(x)+G_{i,-i,-1}(x)\nonumber \\
 & -\frac{1}{2}G_{i,-i,0}(x)+G_{i,i,-1}(x)-\frac{1}{2}G_{i,i,0}(x)-3G_{1,0,0}(x)-4G_{\frac{1}{2}\left(-1-i\sqrt{3}\right),-1,-1}(x)\nonumber \\
 & +2G_{\frac{1}{2}\left(-1-i\sqrt{3}\right),-1,0}(x)-3G_{\frac{1}{2}\left(-1-i\sqrt{3}\right),0,0}(x)+2G_{\frac{1}{2}\left(-1-i\sqrt{3}\right),-i,-1}(x)-G_{\frac{1}{2}\left(-1-i\sqrt{3}\right),-i,0}(x)\nonumber \\
 & +2G_{\frac{1}{2}\left(-1-i\sqrt{3}\right),i,-1}(x)-G_{\frac{1}{2}\left(-1-i\sqrt{3}\right),i,0}(x)-4G_{\frac{1}{2}\left(-1+i\sqrt{3}\right),-1,-1}(x)+2G_{\frac{1}{2}\left(-1+i\sqrt{3}\right),-1,0}(x)\nonumber \\
 & -3G_{\frac{1}{2}\left(-1+i\sqrt{3}\right),0,0}(x)+2G_{\frac{1}{2}\left(-1+i\sqrt{3}\right),-i,-1}(x)-G_{\frac{1}{2}\left(-1+i\sqrt{3}\right),-i,0}(x)+2G_{\frac{1}{2}\left(-1+i\sqrt{3}\right),i,-1}(x)\nonumber \\
 & -G_{\frac{1}{2}\left(-1+i\sqrt{3}\right),i,0}(x)+\frac{1}{12}\left(-6\pi^{2}\log(2)+\log^{3}(2)-15\zeta(3)\right)]+{\cal O}(\epsilon^{4}),\\
 & F_{29}=\epsilon^{2}[4G_{-1,-1}(x)-2G_{-1,0}(x)-2G_{0,-1}(x)+3G_{0,0}(x)-2\log(2)G_{-1}(x)\nonumber \\
 & +\log(2)G_{0}(x)+\frac{1}{4}\left(\pi^{2}+2\log^{2}(2)\right)]\nonumber \\
 & +\epsilon^{3}[-48G_{-1,-1,-1}(x)+24G_{-1,-1,0}(x)+16G_{-1,0,-1}(x)-14G_{-1,0,0}(x)\nonumber \\
 & +8G_{-1,-i,-1}(x)-4G_{-1,-i,0}(x)+8G_{-1,i,-1}(x)-4G_{-1,i,0}(x)\nonumber \\
 & +20G_{0,-1,-1}(x)-22G_{0,-1,0}(x)-8G_{0,0,-1}(x)+9G_{0,0,0}(x)-2G_{0,-i,-1}(x)\nonumber \\
 & +G_{0,-i,0}(x)-2G_{0,i,-1}(x)+G_{0,i,0}(x)-4G_{0,1,0}(x)+6G_{1,0,0}(x)+4G_{\frac{1}{2}\left(-1-i\sqrt{3}\right),-1,-1}(x)\nonumber \\
 & -2G_{\frac{1}{2}\left(-1-i\sqrt{3}\right),-1,0}(x)+3G_{\frac{1}{2}\left(-1-i\sqrt{3}\right),0,0}(x)-2G_{\frac{1}{2}\left(-1-i\sqrt{3}\right),-i,-1}(x)+G_{\frac{1}{2}\left(-1-i\sqrt{3}\right),-i,0}(x)\nonumber \\
 & -2G_{\frac{1}{2}\left(-1-i\sqrt{3}\right),i,-1}(x)+G_{\frac{1}{2}\left(-1-i\sqrt{3}\right),i,0}(x)+4G_{\frac{1}{2}\left(-1+i\sqrt{3}\right),-1,-1}(x)-2G_{\frac{1}{2}\left(-1+i\sqrt{3}\right),-1,0}(x)\nonumber \\
 & +3G_{\frac{1}{2}\left(-1+i\sqrt{3}\right),0,0}(x)-2G_{\frac{1}{2}\left(-1+i\sqrt{3}\right),-i,-1}(x)+G_{\frac{1}{2}\left(-1+i\sqrt{3}\right),-i,0}(x)-2G_{\frac{1}{2}\left(-1+i\sqrt{3}\right),i,-1}(x)\nonumber \\
 & +G_{\frac{1}{2}\left(-1+i\sqrt{3}\right),i,0}(x)+24\log(2)G_{-1,-1}(x)-8\log(2)G_{-1,0}(x)-4\log(2)G_{-1,-i}(x)\nonumber \\
 & -4\log(2)G_{-1,i}(x)-10\log(2)G_{0,-1}(x)+4\log(2)G_{0,0}(x)+\log(2)G_{0,-i}(x)\nonumber \\
 & +\log(2)G_{0,i}(x)-2\log(2)G_{\frac{1}{2}\left(-1-i\sqrt{3}\right),-1}(x)+\log(2)G_{\frac{1}{2}\left(-1-i\sqrt{3}\right),-i}(x)\nonumber \\
 & +\log(2)G_{\frac{1}{2}\left(-1-i\sqrt{3}\right),i}(x)-2\log(2)G_{\frac{1}{2}\left(-1+i\sqrt{3}\right),-1}(x)+\log(2)G_{\frac{1}{2}\left(-1+i\sqrt{3}\right),-i}(x)\nonumber \\
 & +\log(2)G_{\frac{1}{2}\left(-1+i\sqrt{3}\right),i}(x)-\frac{3}{4}\pi^{2}G_{0}(x)+\frac{5}{12}\pi^{2}G_{\frac{1}{2}\left(-1-i\sqrt{3}\right)}(x)+\frac{5}{12}\pi^{2}G_{\frac{1}{2}\left(-1+i\sqrt{3}\right)}(x)\nonumber \\
 & -4\log^{2}(2)G_{-1}(x)+2\log^{2}(2)G_{0}(x)-\frac{33\zeta(3)}{8}+\frac{2\log^{3}(2)}{3}]+{\cal O}(\epsilon^{4}),\\
 & F_{30}=\epsilon^{3}[2i\pi H_{-1,-1}(y)-2H_{-1,-1,0}(y)+\frac{1}{3}\pi^{2}H_{-1}(y)]+{\cal O}(\epsilon^{4}),\\
 & F_{31}=\epsilon^{2}[\frac{1}{2}i\left(\pi H_{-1}(y)+iH_{-1,0}(y)\right)]\nonumber \\
 & +\epsilon^{3}[\frac{5}{2}i\pi H_{-1,-1}(y)-2i\pi H_{-1,0}(y)-i\pi H_{0,-1}(y)-\frac{5}{2}H_{-1,-1,0}(y)+2H_{-1,0,0}(y)\nonumber \\
 & +H_{0,-1,0}(y)-\frac{7}{12}\pi^{2}H_{-1}(y)]+{\cal O}(\epsilon^{4}),\\
 & F_{32}=\epsilon^{3}[4G_{0,0,-1}(x)-4G_{0,-i,-1}(x)+2G_{0,-i,0}(x)-4G_{0,i,-1}(x)+2G_{0,i,0}(x)\nonumber \\
 & -4G_{-i,0,-1}(x)-2G_{-i,0,0}(x)+4G_{-i,-i,-1}(x)-2G_{-i,-i,0}(x)+4G_{-i,i,-1}(x)-2G_{-i,i,0}(x)\nonumber \\
 & -4G_{i,0,-1}(x)-2G_{i,0,0}(x)+4G_{i,-i,-1}(x)-2G_{i,-i,0}(x)+4G_{i,i,-1}(x)-2G_{i,i,0}(x)\nonumber \\
 & +4G_{1,0,0}(x)-2\log(2)G_{0,0}(x)+2\log(2)G_{0,-i}(x)+2\log(2)G_{0,i}(x)+2\log(2)G_{-i,0}(x)\nonumber \\
 & -2\log(2)G_{-i,-i}(x)-2\log(2)G_{-i,i}(x)+2\log(2)G_{i,0}(x)-2\log(2)G_{i,-i}(x)\nonumber \\
 & -2\log(2)G_{i,i}(x)+\frac{1}{6}\pi^{2}G_{0}(x)-\frac{1}{6}\pi^{2}G_{-i}(x)-\frac{1}{6}\pi^{2}G_{i}(x)-\log^{2}(2)G_{0}(x)\nonumber \\
 & +\log^{2}(2)G_{-i}(x)+\log^{2}(2)G_{i}(x)+\frac{1}{12}\left(-3\zeta(3)-4\log^{3}(2)+2\pi^{2}\log(2)\right)]+{\cal O}(\epsilon^{4}),\\
 & F_{33}=\epsilon^{2}[\frac{1}{12}(-12G_{0,-1}(x)+12G_{-i,-1}(x)-6G_{-i,0}(x)+12G_{i,-1}(x)-6G_{i,0}(x)\nonumber \\
 & +\log(64)G_{0}(x)-6\log(2)G_{-i}(x)-6\log(2)G_{i}(x)-\pi^{2}+3\log^{2}(2))]\nonumber \\
 & +\epsilon^{3}[4G_{-1,0,-1}(x)-4G_{-1,-i,-1}(x)+2G_{-1,-i,0}(x)-4G_{-1,i,-1}(x)+2G_{-1,i,0}(x)\nonumber \\
 & +8G_{0,-1,-1}(x)-G_{0,-1,0}(x)-G_{0,0,-1}(x)-3G_{0,-i,-1}(x)+\frac{3}{2}G_{0,-i,0}(x)-3G_{0,i,-1}(x)\nonumber \\
 & +\frac{3}{2}G_{0,i,0}(x)+G_{0,1,0}(x)-8G_{-i,-1,-1}(x)+4G_{-i,-1,0}(x)-G_{-i,0,-1}(x)-\frac{7}{2}G_{-i,0,0}(x)\nonumber \\
 & +5G_{-i,-i,-1}(x)-\frac{5}{2}G_{-i,-i,0}(x)+5G_{-i,i,-1}(x)-\frac{5}{2}G_{-i,i,0}(x)-8G_{i,-1,-1}(x)\nonumber \\
 & +4G_{i,-1,0}(x)-G_{i,0,-1}(x)-\frac{7}{2}G_{i,0,0}(x)+5G_{i,-i,-1}(x)-\frac{5}{2}G_{i,-i,0}(x)\nonumber \\
 & +5G_{i,i,-1}(x)-\frac{5}{2}G_{i,i,0}(x)+3G_{1,0,0}(x)-2\log(2)G_{-1,0}(x)+2\log(2)G_{-1,-i}(x)\nonumber \\
 & +2\log(2)G_{-1,i}(x)-4\log(2)G_{0,-1}(x)+\frac{1}{2}\log(2)G_{0,0}(x)+\frac{3}{2}\log(2)G_{0,-i}(x)\nonumber \\
 & +\frac{3}{2}\log(2)G_{0,i}(x)+4\log(2)G_{-i,-1}(x)+\frac{1}{2}\log(2)G_{-i,0}(x)-\frac{5}{2}\log(2)G_{-i,-i}(x)\nonumber \\
 & -\frac{5}{2}\log(2)G_{-i,i}(x)+4\log(2)G_{i,-1}(x)+\frac{1}{2}\log(2)G_{i,0}(x)-\frac{5}{2}\log(2)G_{i,-i}(x)\nonumber \\
 & -\frac{5}{2}\log(2)G_{i,i}(x)+\frac{1}{3}\pi^{2}G_{-1}(x)-\frac{1}{12}\pi^{2}G_{-i}(x)-\frac{1}{12}\pi^{2}G_{i}(x)-\log^{2}(2)G_{-1}(x)\nonumber \\
 & +\frac{1}{4}\log^{2}(2)G_{0}(x)+\frac{1}{4}\log^{2}(2)G_{-i}(x)+\frac{1}{4}\log^{2}(2)G_{i}(x)\nonumber \\
 & +\frac{1}{12}\left(-18\zeta(3)+\log^{3}(2)-\pi^{2}\log(2)\right)]+{\cal O}(\epsilon^{4}),\\
 & F_{34}=\epsilon^{2}[H_{-1,-1}(z)+2H_{-1,1}(z)-H_{0,1}(z)+\frac{1}{2}(\log(4)-2i\pi)H_{-1}(z)-\frac{\log^{2}(2)}{2}+i\pi\log(2)]\nonumber \\
 & +\epsilon^{3}[5i\pi H_{-1,-1}(z)+2i\pi H_{1,-1}(z)-5H_{-1,-1,-1}(z)-6H_{-1,-1,1}(z)+6H_{-1,0,1}(z)\nonumber \\
 & +8H_{-1,1,1}(z)-H_{0,0,1}(z)-4H_{0,1,1}(z)-2H_{1,-1,-1}(z)-4H_{1,-1,1}(z)+2H_{1,0,1}(z)\nonumber \\
 & -\log(8)H_{-1,-1}(z)-\log(4)H_{-1,-1}(z)-\log(4)H_{1,-1}(z)\nonumber \\
 & +\frac{1}{6}\left(8\pi^{2}+3\log^{2}(2)-6i\pi\log(2)\right)H_{-1}(z)+\log^{2}(2)H_{1}(z)-2i\pi\log(2)H_{1}(z)\nonumber \\
 & -\frac{\zeta(3)}{4}+\frac{i\pi^{3}}{6}+\frac{5\log^{3}(2)}{6}-\frac{5}{2}i\pi\log^{2}(2)-\frac{4}{3}\pi^{2}\log(2)]+{\cal O}(\epsilon^{4}),\\
 & F_{35}=\epsilon^{3}[-H_{-1,0,1}(z)-H_{0,-1,-1}(z)-H_{0,-1,1}(z)+H_{0,0,1}(z)\nonumber \\
 & +i(\pi+i\log(2))H_{0,-1}(z)+\frac{1}{6}\pi^{2}H_{-1}(z)+\frac{\zeta(3)}{8}-\frac{i\pi^{3}}{12}-\frac{1}{6}\pi^{2}\log(2)]+{\cal O}(\epsilon^{4}),\\
 & F_{36}=\epsilon^{2}[\frac{1}{4}(-2H_{-1,-1}(z)-4H_{-1,1}(z)+2H_{0,1}(z)+2i(\pi+i\log(2))H_{-1}(z)\nonumber \\
 & +\log^{2}(2)-2i\pi\log(2))]\nonumber \\
 & +\epsilon^{3}[\frac{1}{12}(-30i\pi H_{-1,-1}(z)-12i\pi H_{1,-1}(z)+30H_{-1,-1,-1}(z)+36H_{-1,-1,1}(z)\nonumber \\
 & -48H_{-1,0,1}(z)-48H_{-1,1,1}(z)-12H_{0,-1,-1}(z)-12H_{0,-1,1}(z)+18H_{0,0,1}(z)\nonumber \\
 & +24H_{0,1,1}(z)+12H_{1,-1,-1}(z)+24H_{1,-1,1}(z)-12H_{1,0,1}(z)+30\log(2)H_{-1,-1}(z)\nonumber \\
 & +12i(\pi+i\log(2))H_{0,-1}(z)+12\log(2)H_{1,-1}(z)-3\left(2\pi^{2}+\log^{2}(2)-2i\pi\log(2)\right)H_{-1}(z)\nonumber \\
 & -6\log^{2}(2)H_{1}(z)+12i\pi\log(2)H_{1}(z)+3\zeta(3)-2i\pi^{3}\nonumber \\
 & -5\log^{3}(2)+15i\pi\log^{2}(2)+\pi^{2}\log(64))]+{\cal O}(\epsilon^{4}),\\
 & F_{37}=\epsilon[\frac{H_{1}(z)}{2}]\nonumber \\
 & +\epsilon^{2}[\frac{1}{2}\left(H_{-1,-1}(z)+H_{-1,1}(z)+4H_{1,1}(z)+(\log(2)-i\pi)H_{-1}(z)\right)]\nonumber \\
 & +\epsilon^{3}[\frac{1}{12}(6(i\pi H_{1,-1}(z)-5H_{-1,-1,-1}(z)-5H_{-1,-1,1}(z)+4H_{-1,0,1}(z)+4H_{-1,1,1}(z)\nonumber \\
 & -H_{1,-1,-1}(z)-2H_{1,-1,1}(z)+7H_{1,0,1}(z)+16H_{1,1,1}(z)+5i(\pi+i\log(2))H_{-1,-1}(z)\nonumber \\
 & -\log(2)H_{1,-1}(z))+9\pi^{2}H_{-1}(z)+(2\pi^{2}+3\log^{2}(2)-6i\pi\log(2))H_{1}(z))]+{\cal O}(\epsilon^{4}),\\
 & F_{38}=\epsilon^{3}[-G_{-3,-1,0}(x)+\frac{1}{2}G_{-3,1,0}(x)+G_{-\frac{1}{3},-1,0}(x)-\frac{1}{2}G_{-\frac{1}{3},0,0}(x)-\frac{1}{2}G_{-\frac{1}{3},1,0}(x)\nonumber \\
 & -2G_{0,-1,0}(x)+\frac{1}{2}G_{0,0,0}(x)+G_{0,1,0}(x)+\frac{1}{2}G_{1,0,0}(x)-\frac{2}{3}\pi^{2}G_{-3}(x)-\frac{1}{3}\pi^{2}G_{-\frac{1}{3}}(x)\nonumber \\
 & +\frac{1}{6}\pi^{2}G_{0}(x)+\frac{1}{6}\left(5\zeta(3)+6\pi^{2}\log(2)-4\pi^{2}\log(3)\right)]+{\cal O}(\epsilon^{4}),\\
 & F_{39}=\epsilon^{2}[-\frac{1}{4}G_{0,0}(x)-\frac{\pi^{2}}{8}]\nonumber \\
 & +\epsilon^{3}[-G_{-3,-1,0}(x)+\frac{1}{2} G_{-3,1,0}(x)+G_{-1,0,0}(x)+G_{-\frac{1}{3},-1,0}(x)\nonumber \\
 &-\frac{1}{2}G_{-\frac{1}{3},0,0}(x)-\frac{1}{2} G_{-\frac{1}{3},1,0}(x)-\frac{1}{2} G_{0,-1,0}(x)-\frac{1}{4}
   G_{0,0,0}(x)\nonumber \\
 &+\frac{1}{2} G_{0,1,0}(x)-\frac{2}{3} \pi ^2 G_{-3}(x)+\frac{1}{2} \pi ^2
   G_{-1}(x)-\frac{1}{3} \pi ^2 G_{-\frac{1}{3}}(x)+\frac{1}{8} \pi ^2 G_0(x)\nonumber \\
 &+\frac{1}{24} \left(-37 \zeta
   (3)+18 \pi ^2 \log (2)-16 \pi ^2 \log (3)\right)]+{\cal O}(\epsilon^{4}),\\
 & F_{40}=\epsilon^{2}[2G_{-1,0}(x)-G_{0,0}(x)-G_{1,0}(x)+\frac{\pi^{2}}{12}]\nonumber \\
 & +\epsilon^{3}[2G_{-3,-1,0}(x)-G_{-3,1,0}(x)-16G_{-1,-1,0}(x)+7G_{-1,0,0}(x)+6G_{-1,1,0}(x)\nonumber \\
 & +10G_{0,-1,0}(x)-4G_{0,0,0}(x)-4G_{0,1,0}(x)+4G_{1,-1,0}(x)-3G_{1,0,0}(x)-G_{1,1,0}(x)\nonumber \\
 & +\frac{4}{3}\pi^{2}G_{-3}(x)-\frac{4}{3}\pi^{2}G_{-1}(x)-\frac{1}{2}\pi^{2}G_{0}(x)+\frac{1}{2}\pi^{2}G_{1}(x)+\frac{67\zeta(3)}{12}\nonumber \\
 & +\frac{4}{3}\pi^{2}\log(3)-\frac{3}{2}\pi^{2}\log(2)]+{\cal O}(\epsilon^{4}),\\
 & F_{41}=\epsilon^{3}[G_{0,0,0}(x)-2G_{1,0,0}(x)+\frac{3\zeta(3)}{2}-\pi^{2}\log(2)]+{\cal O}(\epsilon^{4}),\\
 & F_{42}=\epsilon^{3}[G_{0,0,0}(x)-2G_{-i,0,0}(x)-2G_{i,0,0}(x)+2G_{1,0,0}(x)+\frac{1}{4}\pi^{2}G_{0}(x)-\frac{1}{4}\pi^{2}G_{-i}(x)\nonumber \\
 & -\frac{1}{4}\pi^{2}G_{i}(x)+\frac{1}{8}\left(\pi^{2}\log(4)-9\zeta(3)\right)]+{\cal O}(\epsilon^{4}),\\
 & F_{43}=\epsilon^{2}[-\frac{1}{4}G_{0,0}(x)-\frac{\pi^{2}}{8}]\nonumber \\
 & +\epsilon^{3}[G_{-1,0,0}(x)+\frac{3}{2}G_{0,-1,0}(x)-\frac{1}{4}G_{0,0,0}(x)+\frac{1}{2}G_{0,1,0}(x)-2G_{-i,0,0}(x)-2G_{i,0,0}(x)\nonumber \\
 & +\frac{3}{2}G_{1,0,0}(x)+\frac{1}{2}\pi^{2}G_{-1}(x)+\frac{1}{24}\pi^{2}G_{0}(x)-\frac{1}{4}\pi^{2}G_{-i}(x)-\frac{1}{4}\pi^{2}G_{i}(x)-2\zeta(3)]+{\cal O}(\epsilon^{4}),\\
 & F_{44}=\epsilon[\frac{1}{4}\left(\log(2)-H_{-1}(z)\right)]\nonumber \\
 & +\epsilon^{2}[\frac{1}{48}(12H_{-1,-1}(z)-12H_{1,-1}(z)+12(\log(8)-2i\pi)H_{-1}(z)+12\log(2)H_{1}(z)\nonumber \\
 & -\pi^{2}-12\log(2)\log(8)+24i\pi\log(2))]\nonumber \\
 & +\epsilon^{3}[\frac{1}{48}(2(-6H_{-1,-1,-1}(z)+6H_{-1,1,-1}(z)+6H_{1,-1,-1}(z)-6H_{1,1,-1}(z)\nonumber \\
 & +(-6\log(8)+12i\pi)H_{-1,-1}(z)-6\log(2)H_{-1,1}(z)+6(\log(8)-2i\pi)H_{1,-1}(z)\nonumber \\
 & +\log(64)H_{1,1}(z)+6\zeta(3)-i\pi^{3}+28\log^{3}(2)-36i\pi\log^{2}(2)-13\pi^{2}\log(2))\nonumber \\
 & +\left(29\pi^{2}+72i\pi\log(2)-12\log(2)\log(32)\right)H_{-1}(z)\nonumber \\
 & -\left(\pi^{2}-24i\pi\log(2)+12\log(2)\log(8)\right)H_{1}(z))]+{\cal O}(\epsilon^{4}),\\
 & F_{45}=\epsilon[\frac{1}{4}\left(\log(2)-H_{-1}(z)\right)]\nonumber \\
 & +\epsilon^{2}[\frac{1}{48}(24H_{-1,-1}(z)-12H_{1,-1}(z)+24(\log(2)-i\pi)H_{-1}(z)+12\log(2)H_{1}(z)-\pi^{2}\nonumber \\
 & -\log(32)\log(64)+24i\pi\log(2))]\nonumber \\
 & +\epsilon^{3}[\frac{1}{96}(96i\pi H_{-1,-1}(z)-48i\pi H_{1,-1}(z)-96H_{-1,-1,-1}(z)+48H_{1,-1,-1}(z)\nonumber \\
 & -24H_{1,1,-1}(z)-96\log(2)H_{-1,-1}(z)+48\log(2)H_{1,-1}(z)+24\log(2)H_{1,1}(z)\nonumber \\
 & +8\left(7\pi^{2}-6\log^{2}(2)+12i\pi\log(2)\right)H_{-1}(z)-2\left(\pi^{2}-24i\pi\log(2)+\log(32)\log(64)\right)H_{1}(z)\nonumber \\
 & +15\zeta(3)-4i\pi^{3}+76\log^{3}(2)-120i\pi\log^{2}(2)-50\pi^{2}\log(2))]+{\cal O}(\epsilon^{4}),\\
 & F_{46}={\cal O}(\epsilon^{4}),\\
 & F_{47}=\epsilon^{3}[i\pi H_{-1,-1}(y)-2i\pi H_{0,-1}(y)-i\pi H_{1,0}(y)-H_{-1,-1,0}(y)+2H_{0,-1,0}(y)\nonumber \\
 & +H_{1,0,0}(y)+\frac{1}{6}\pi^{2}H_{-1}(y)]+{\cal O}(\epsilon^{4}),\\
 & F_{48}={\cal O}(\epsilon^{4}),\\
 & F_{49}={\cal O}(\epsilon^{4}),\\
 & F_{50}=\epsilon^{2}[-2G_{0,-1}(x)+2G_{-i,-1}(x)-G_{-i,0}(x)+2G_{i,-1}(x)-G_{i,0}(x)+\log(2)G_{0}(x)\nonumber \\
 & -\log(2)G_{-i}(x)-\log(2)G_{i}(x)+\frac{1}{6}\left(3\log^{2}(2)-\pi^{2}\right)]\nonumber \\
 & +\epsilon^{3}[-2\log^{2}(2)G_{-1}(x)+\frac{2}{3}\pi^{2}G_{-1}(x)+2\log^{2}(2)G_{0}(x)-\frac{1}{12}\pi^{2}G_{0}(x)\nonumber \\
 & -\log^{2}(2)G_{-i}(x)+\frac{1}{3}\pi^{2}G_{-i}(x)-\log^{2}(2)G_{i}(x)+\frac{1}{3}\pi^{2}G_{i}(x)\nonumber \\
 & -\frac{5}{12}\pi^{2}G_{\frac{1}{2}\left(-1-i\sqrt{3}\right)}(x)-\frac{5}{12}\pi^{2}G_{\frac{1}{2}\left(-1+i\sqrt{3}\right)}(x)\nonumber \\
 & -4\log(2)G_{-1,0}(x)+4\log(2)G_{-1,-i}(x)+4\log(2)G_{-1,i}(x)-10\log(2)G_{0,-1}(x)\nonumber \\
 & +2\log(2)G_{0,0}(x)+\log(2)G_{0,-i}(x)+\log(2)G_{0,i}(x)+8\log(2)G_{-i,-1}(x)\nonumber \\
 & -2\log(2)G_{-i,0}(x)-2\log(2)G_{-i,-i}(x)-2\log(2)G_{-i,i}(x)+8\log(2)G_{i,-1}(x)\nonumber \\
 & -2\log(2)G_{i,0}(x)-2\log(2)G_{i,-i}(x)-2\log(2)G_{i,i}(x)+2\log(2)G_{\frac{1}{2}\left(-1-i\sqrt{3}\right),-1}(x)\nonumber \\
 & -\log(2)G_{\frac{1}{2}\left(-1-i\sqrt{3}\right),-i}(x)-\log(2)G_{\frac{1}{2}\left(-1-i\sqrt{3}\right),i}(x)+2\log(2)G_{\frac{1}{2}\left(-1+i\sqrt{3}\right),-1}(x)\nonumber \\
 & -\log(2)G_{\frac{1}{2}\left(-1+i\sqrt{3}\right),-i}(x)-\log(2)G_{\frac{1}{2}\left(-1+i\sqrt{3}\right),i}(x)+8G_{-1,0,-1}(x)-8G_{-1,-i,-1}(x)\nonumber \\
 & +4G_{-1,-i,0}(x)-8G_{-1,i,-1}(x)+4G_{-1,i,0}(x)+20G_{0,-1,-1}(x)-4G_{0,-1,0}(x)-4G_{0,0,-1}(x)\nonumber \\
 & -2G_{0,-i,-1}(x)+G_{0,-i,0}(x)-2G_{0,i,-1}(x)+G_{0,i,0}(x)-16G_{-i,-1,-1}(x)+8G_{-i,-1,0}(x)\nonumber \\
 & +4G_{-i,0,-1}(x)-2G_{-i,0,0}(x)+4G_{-i,-i,-1}(x)-2G_{-i,-i,0}(x)+4G_{-i,i,-1}(x)-2G_{-i,i,0}(x)\nonumber \\
 & -16G_{i,-1,-1}(x)+8G_{i,-1,0}(x)+4G_{i,0,-1}(x)-2G_{i,0,0}(x)+4G_{i,-i,-1}(x)\nonumber \\
 & -2G_{i,-i,0}(x)+4G_{i,i,-1}(x)-2G_{i,i,0}(x)-4G_{\frac{1}{2}\left(-1-i\sqrt{3}\right),-1,-1}(x)+2G_{\frac{1}{2}\left(-1-i\sqrt{3}\right),-1,0}(x)\nonumber \\
 & -3G_{\frac{1}{2}\left(-1-i\sqrt{3}\right),0,0}(x)+2G_{\frac{1}{2}\left(-1-i\sqrt{3}\right),-i,-1}(x)-G_{\frac{1}{2}\left(-1-i\sqrt{3}\right),-i,0}(x)+2G_{\frac{1}{2}\left(-1-i\sqrt{3}\right),i,-1}(x)\nonumber \\
 & -G_{\frac{1}{2}\left(-1-i\sqrt{3}\right),i,0}(x)-4G_{\frac{1}{2}\left(-1+i\sqrt{3}\right),-1,-1}(x)+2G_{\frac{1}{2}\left(-1+i\sqrt{3}\right),-1,0}(x)-3G_{\frac{1}{2}\left(-1+i\sqrt{3}\right),0,0}(x)\nonumber \\
 & +2G_{\frac{1}{2}\left(-1+i\sqrt{3}\right),-i,-1}(x)-G_{\frac{1}{2}\left(-1+i\sqrt{3}\right),-i,0}(x)+2G_{\frac{1}{2}\left(-1+i\sqrt{3}\right),i,-1}(x)-G_{\frac{1}{2}\left(-1+i\sqrt{3}\right),i,0}(x)\nonumber \\
 & +\frac{1}{24}\left(-16\pi^{2}\log(2)+16\log^{3}(2)-75\zeta(3)\right)]+{\cal O}(\epsilon^{4}),\\
 & F_{51}=\epsilon^{3}[-G_{0,-1,0}(x)-G_{0,0,-1}(x)-\frac{1}{2}G_{0,1,0}(x)+2G_{-i,-1,0}(x)+2G_{-i,0,-1}(x)\nonumber \\
 & -2G_{-i,0,0}(x)+2G_{i,-1,0}(x)+2G_{i,0,-1}(x)-2G_{i,0,0}(x)-2G_{1,-1,0}(x)-2G_{1,0,-1}(x)\nonumber \\
 & +\frac{3}{2}G_{1,0,0}(x)+G_{1,1,0}(x)+\frac{1}{4}\log(8)G_{0,0}(x)-\frac{1}{4}\log(2)G_{0,0}(x)-\log(2)G_{-i,0}(x)\nonumber \\
 & -\log(2)G_{i,0}(x)+\log(2)G_{1,0}(x)-\frac{1}{24}\pi^{2}G_{0}(x)-\frac{1}{6}\pi^{2}G_{1}(x)\nonumber \\
 & +\frac{1}{48}\left(3\zeta(3)+8\pi^{2}\log(2)\right)]+{\cal O}(\epsilon^{4}),\\
 & F_{52}=\frac{1}{4}+\epsilon[-2G_{-1}(x)+G_{0}(x)+\log(2)]\nonumber \\
 & +\epsilon^{2}[16G_{-1,-1}(x)-8G_{-1,0}(x)-4G_{0,-1}(x)+5G_{0,0}(x)-4G_{-i,-1}(x)+2G_{-i,0}(x)\nonumber \\
 & -4G_{i,-1}(x)+2G_{i,0}(x)-8\log(2)G_{-1}(x)+2\log(2)G_{0}(x)+2\log(2)G_{-i}(x)\nonumber \\
 & +2\log(2)G_{i}(x)+\frac{\pi^{2}}{6}+\log^{2}(2)]\nonumber \\
 & +\epsilon^{3}[-9\log^{2}(2)G_{-1}(x)+\log^{2}(2)G_{0}(x)-\frac{3}{4}\pi^{2}G_{0}(x)+3\log^{2}(2)G_{-i}(x)-\frac{1}{6}\pi^{2}G_{-i}(x)\nonumber \\
 & +3\log^{2}(2)G_{i}(x)-\frac{1}{6}\pi^{2}G_{i}(x)+\log^{2}(2)G_{1}(x)+\frac{1}{6}\pi^{2}G_{1}(x)+\frac{5}{6}\pi^{2}G_{\frac{1}{2}\left(-1-i\sqrt{3}\right)}(x)\nonumber \\
 & +\frac{5}{6}\pi^{2}G_{\frac{1}{2}\left(-1+i\sqrt{3}\right)}(x)+64\log(2)G_{-1,-1}(x)-2\log(8)G_{-1,0}(x)-12\log(2)G_{-1,0}(x)\nonumber \\
 & -14\log(2)G_{-1,-i}(x)-14\log(2)G_{-1,i}(x)-8\log(2)G_{0,-1}(x)+\frac{3}{2}\log(8)G_{0,0}(x)\nonumber \\
 & +\frac{1}{2}\log(2)G_{0,0}(x)+2\log(2)G_{0,-i}(x)+2\log(2)G_{0,i}(x)-20\log(2)G_{-i,-1}(x)\nonumber \\
 & +6\log(2)G_{-i,0}(x)+4\log(2)G_{-i,-i}(x)+4\log(2)G_{-i,i}(x)-20\log(2)G_{i,-1}(x)\nonumber \\
 & +6\log(2)G_{i,0}(x)+4\log(2)G_{i,-i}(x)+4\log(2)G_{i,i}(x)-\log(8)G_{1,0}(x)+5\log(2)G_{1,0}(x)\nonumber \\
 & -2\log(2)G_{1,-i}(x)-2\log(2)G_{1,i}(x)-4\log(2)G_{\frac{1}{2}\left(-1-i\sqrt{3}\right),-1}(x)\nonumber \\
 & +2\log(2)G_{\frac{1}{2}\left(-1-i\sqrt{3}\right),-i}(x)+2\log(2)G_{\frac{1}{2}\left(-1-i\sqrt{3}\right),i}(x)-4\log(2)G_{\frac{1}{2}\left(-1+i\sqrt{3}\right),-1}(x)\nonumber \\
 & +2\log(2)G_{\frac{1}{2}\left(-1+i\sqrt{3}\right),-i}(x)+2\log(2)G_{\frac{1}{2}\left(-1+i\sqrt{3}\right),i}(x)-128G_{-1,-1,-1}(x)\nonumber \\
 & +64G_{-1,-1,0}(x)+36G_{-1,0,-1}(x)-30G_{-1,0,0}(x)+28G_{-1,-i,-1}(x)-14G_{-1,-i,0}(x)\nonumber \\
 & +28G_{-1,i,-1}(x)-14G_{-1,i,0}(x)+16G_{0,-1,-1}(x)-28G_{0,-1,0}(x)-10G_{0,0,-1}(x)\nonumber \\
 & +13G_{0,0,0}(x)-4G_{0,-i,-1}(x)+2G_{0,-i,0}(x)-4G_{0,i,-1}(x)+2G_{0,i,0}(x)-7G_{0,1,0}(x)\nonumber \\
 & +40G_{-i,-1,-1}(x)-20G_{-i,-1,0}(x)-12G_{-i,0,-1}(x)+10G_{-i,0,0}(x)-8G_{-i,-i,-1}(x)\nonumber \\
 & +4G_{-i,-i,0}(x)-8G_{-i,i,-1}(x)+4G_{-i,i,0}(x)+40G_{i,-1,-1}(x)-20G_{i,-1,0}(x)\nonumber \\
 & -12G_{i,0,-1}(x)+10G_{i,0,0}(x)-8G_{i,-i,-1}(x)+4G_{i,-i,0}(x)-8G_{i,i,-1}(x)+4G_{i,i,0}(x)\nonumber \\
 & -4G_{1,0,-1}(x)+5G_{1,0,0}(x)+4G_{1,-i,-1}(x)-2G_{1,-i,0}(x)+4G_{1,i,-1}(x)-2G_{1,i,0}(x)\nonumber \\
 & +8G_{\frac{1}{2}\left(-1-i\sqrt{3}\right),-1,-1}(x)-4G_{\frac{1}{2}\left(-1-i\sqrt{3}\right),-1,0}(x)+6G_{\frac{1}{2}\left(-1-i\sqrt{3}\right),0,0}(x)\nonumber \\
 & -4G_{\frac{1}{2}\left(-1-i\sqrt{3}\right),-i,-1}(x)+2G_{\frac{1}{2}\left(-1-i\sqrt{3}\right),-i,0}(x)\nonumber \\
 & -4G_{\frac{1}{2}\left(-1-i\sqrt{3}\right),i,-1}(x)+2G_{\frac{1}{2}\left(-1-i\sqrt{3}\right),i,0}(x)+8G_{\frac{1}{2}\left(-1+i\sqrt{3}\right),-1,-1}(x)\nonumber \\
 & -4G_{\frac{1}{2}\left(-1+i\sqrt{3}\right),-1,0}(x)+6G_{\frac{1}{2}\left(-1+i\sqrt{3}\right),0,0}(x)-4G_{\frac{1}{2}\left(-1+i\sqrt{3}\right),-i,-1}(x)\nonumber \\
 & +2G_{\frac{1}{2}\left(-1+i\sqrt{3}\right),-i,0}(x)-4G_{\frac{1}{2}\left(-1+i\sqrt{3}\right),i,-1}(x)+2G_{\frac{1}{2}\left(-1+i\sqrt{3}\right),i,0}(x)\nonumber \\
 & +\frac{1}{24}\left(8\pi^{2}\log(2)+8\log^{3}(2)+63\zeta(3)\right)]+{\cal O}(\epsilon^{4}),\\
 & F_{53}={\cal O}(\epsilon^{4}),\\
 & F_{54}=\epsilon^{3}[-G_{0,-1,0}(x)-3G_{0,0,-1}(x)+2G_{0,-i,-1}(x)-G_{0,-i,0}(x)+2G_{0,i,-1}(x)\nonumber \\
 & -G_{0,i,0}(x)+\frac{1}{2}G_{0,1,0}(x)+2G_{1,-1,0}(x)+2G_{1,0,-1}(x)-\frac{3}{2}G_{1,0,0}(x)-G_{1,1,0}(x)\nonumber \\
 & +\frac{3}{2}\log(2)G_{0,0}(x)-\log(2)G_{0,-i}(x)-\log(2)G_{0,i}(x)-\log(2)G_{1,0}(x)-\frac{1}{8}\pi^{2}G_{0}(x)\nonumber \\
 & +\frac{1}{6}\pi^{2}G_{1}(x)+\frac{1}{2}\log^{2}(2)G_{0}(x)+\frac{1}{48}\left(-3\zeta(3)-8\pi^{2}\log(2)\right)]+O(\epsilon^{4}),\\
 & F_{55}=\epsilon^{3}[G_{-1,0,-1}(x)+\frac{3}{2}G_{-1,0,0}(x)-G_{-1,-i,-1}(x)+\frac{1}{2}G_{-1,-i,0}(x)-G_{-1,i,-1}(x)\nonumber \\
 & +\frac{1}{2}G_{-1,i,0}(x)-\frac{1}{2}G_{0,-1,0}(x)+\frac{3}{2}G_{0,0,-1}(x)-2G_{0,-i,-1}(x)+G_{0,-i,0}(x)-2G_{0,i,-1}(x)\nonumber \\
 & +G_{0,i,0}(x)+\frac{1}{4}G_{0,1,0}(x)-2G_{-i,0,-1}(x)-G_{-i,0,0}(x)+2G_{-i,-i,-1}(x)-G_{-i,-i,0}(x)\nonumber \\
 & +2G_{-i,i,-1}(x)-G_{-i,i,0}(x)-2G_{i,0,-1}(x)-G_{i,0,0}(x)+2G_{i,-i,-1}(x)-G_{i,-i,0}(x)\nonumber \\
 & +2G_{i,i,-1}(x)-G_{i,i,0}(x)-G_{1,0,-1}(x)+\frac{5}{4}G_{1,0,0}(x)+G_{1,-i,-1}(x)-\frac{1}{2}G_{1,-i,0}(x)\nonumber \\
 & +G_{1,i,-1}(x)-\frac{1}{2}G_{1,i,0}(x)-\frac{1}{2}\log(2)G_{-1,0}(x)+\frac{1}{2}\log(2)G_{-1,-i}(x)\nonumber \\
 & +\frac{1}{2}\log(2)G_{-1,i}(x)-\frac{3}{4}\log(2)G_{0,0}(x)+\log(2)G_{0,-i}(x)+\log(2)G_{0,i}(x)\nonumber \\
 & +\log(2)G_{-i,0}(x)-\log(2)G_{-i,-i}(x)-\log(2)G_{-i,i}(x)+\log(2)G_{i,0}(x)-\log(2)G_{i,-i}(x)\nonumber \\
 & -\log(2)G_{i,i}(x)+\frac{1}{2}\log(2)G_{1,0}(x)-\frac{1}{2}\log(2)G_{1,-i}(x)-\frac{1}{2}\log(2)G_{1,i}(x)\nonumber \\
 & -\frac{1}{6}\pi^{2}G_{-1}(x)+\frac{5}{48}\pi^{2}G_{0}(x)-\frac{1}{12}\pi^{2}G_{-i}(x)-\frac{1}{12}\pi^{2}G_{i}(x)+\frac{1}{24}\pi^{2}G_{1}(x)\nonumber \\
 & -\frac{1}{4}\log^{2}(2)G_{-1}(x)-\frac{1}{2}\log^{2}(2)G_{0}(x)+\frac{1}{2}\log^{2}(2)G_{-i}(x)+\frac{1}{2}\log^{2}(2)G_{i}(x)\nonumber \\
 & +\frac{1}{4}\log^{2}(2)G_{1}(x)+\frac{1}{96}\left(-129\zeta(3)-16\log^{3}(2)+20\pi^{2}\log(2)\right)]+{\cal O}(\epsilon^{4}),\\
 & F_{56}={\cal O}(\epsilon^{4}),\\
 & F_{57}={\cal O}(\epsilon^{4}),\\
 & F_{58}=\epsilon^{3}[-\frac{1}{2}G_{-3,-1,0}(x)+\frac{1}{4}G_{-3,1,0}(x)-\frac{1}{2}G_{-\frac{1}{3},-1,0}(x)+\frac{1}{4}G_{-\frac{1}{3},0,0}(x)+\frac{1}{4}G_{-\frac{1}{3},1,0}(x)\nonumber \\
 & -\frac{3}{4}G_{0,0,0}(x)+G_{1,-1,0}(x)-\frac{1}{4}G_{1,0,0}(x)-\frac{1}{2}G_{1,1,0}(x)-\frac{1}{3}\pi^{2}G_{-3}(x)+\frac{1}{6}\pi^{2}G_{-\frac{1}{3}}(x)\nonumber \\
 & -\frac{1}{4}\pi^{2}G_{0}(x)+\frac{1}{6}\pi^{2}G_{1}(x)+\frac{1}{6}\left(7\zeta(3)-2\pi^{2}\log(3)\right)]+{\cal O}(\epsilon^{4}),\\
 & F_{59}=\epsilon^{3}[\frac{1}{4}G_{-3,-1,0}(x)-\frac{1}{8}G_{-3,1,0}(x)+G_{-1,0,0}(x)-\frac{1}{4}G_{-\frac{1}{3},-1,0}(x)+\frac{1}{8}G_{-\frac{1}{3},0,0}(x)\nonumber \\
 & +\frac{1}{8}G_{-\frac{1}{3},1,0}(x)-\frac{1}{2}G_{0,-1,0}(x)+\frac{3}{8}G_{0,0,0}(x)+\frac{1}{4}G_{0,1,0}(x)-G_{-i,0,0}(x)-G_{i,0,0}(x)\nonumber \\
 & +\frac{3}{8}G_{1,0,0}(x)+\frac{1}{6}\pi^{2}G_{-3}(x)-\frac{1}{2}\pi^{2}G_{-1}(x)+\frac{1}{12}\pi^{2}G_{-\frac{1}{3}}(x)+\frac{1}{6}\pi^{2}G_{0}(x)\nonumber \\
 & -\frac{1}{8}\pi^{2}G_{-i}(x)-\frac{1}{8}\pi^{2}G_{i}(x)+\frac{1}{48}\left(-91\zeta(3)+6\pi^{2}\log(2)+8\pi^{2}\log(3)\right)]+{\cal O}(\epsilon^{4}),\\
 & F_{60}={\cal O}(\epsilon^{4}),\\
 & F_{61}=\epsilon^{3}[\frac{1}{2}G_{-3,-1,0}(x)-\frac{1}{4}G_{-3,1,0}(x)+\frac{1}{2}G_{-\frac{1}{3},-1,0}(x)-\frac{1}{4}G_{-\frac{1}{3},0,0}(x)-\frac{1}{4}G_{-\frac{1}{3},1,0}(x)\nonumber \\
 & -\frac{3}{4}G_{0,0,0}(x)-G_{1,-1,0}(x)+\frac{1}{4}G_{1,0,0}(x)+\frac{1}{2}G_{1,1,0}(x)+\frac{1}{3}\pi^{2}G_{-3}(x)\nonumber \\
 & -\frac{1}{6}\pi^{2}G_{-\frac{1}{3}}(x)-\frac{1}{6}\pi^{2}G_{1}(x)+\frac{1}{6}\left(2\pi^{2}\log(3)-7\zeta(3)\right)]+{\cal O}(\epsilon^{4}),\\
 & F_{62}=\epsilon^{3}[\frac{1}{24}(-24H_{1,-1,-1}(z)-48H_{1,-1,1}(z)+24H_{1,0,1}(z)+(-12\log(4)\nonumber \\
 & +24i\pi)H_{1,-1}(z)+12\log^{2}(2)H_{1}(z)-24i\pi\log(2)H_{1}(z)-33\zeta(3)+2i\pi^{3}\nonumber \\
 & +4\log^{3}(2)-12i\pi\log^{2}(2)+\pi^{2}\log(16))]+{\cal O}(\epsilon^{4}),\\
 & F_{63}={\cal O}(\epsilon^{4}),\\
 & F_{64}=\epsilon[\frac{1}{12}\left(\log(2)-H_{-1}(z)\right)]\nonumber \\
 & +\epsilon^{2}[\frac{1}{72}(6H_{-1,-1}(z)+12H_{-1,1}(z)-12H_{1,-1}(z)+6(\log(32)-3i\pi)H_{-1}(z)\nonumber \\
 & +12\log(2)H_{1}(z)-2\pi^{2}-21\log^{2}(2)+18i\pi\log(2))]\nonumber \\
 & +\epsilon^{3}[\frac{1}{72}(18i\pi H_{-1,-1}(z)-6H_{-1,-1,-1}(z)-12H_{-1,-1,1}(z)+36H_{-1,0,1}(z)+12H_{-1,1,-1}(z)\nonumber \\
 & +48H_{-1,1,1}(z)-24H_{1,-1,-1}(z)-48H_{1,-1,1}(z)+36H_{1,0,1}(z)-24H_{1,1,-1}(z)\nonumber \\
 & +30\log(32)H_{-1,-1}(z)-72\log(8)H_{-1,-1}(z)+36\log(2)H_{-1,-1}(z)-12\log(2)H_{-1,1}(z)\nonumber \\
 & +12\log(32)H_{1,-1}(z)-36\log(2)H_{1,-1}(z)+24\log(2)H_{1,1}(z)-4\left(\pi^{2}+6\log^{2}(2)\right)H_{1}(z)\nonumber \\
 & +3\left(7\pi^{2}+18i\pi\log(2)-\log(2)\log(131072)\right)H_{-1}(z)-3\zeta(3)+37\log^{3}(2)\nonumber \\
 & -63i\pi\log^{2}(2)-17\pi^{2}\log(2))]+{\cal O}(\epsilon^{4}),\\
 & F_{65}={\cal O}(\epsilon^{4}),\\
 & F_{66}=\epsilon^{3}[-G_{-3,-1,0}(x)+\frac{1}{2}G_{-3,1,0}(x)-G_{-\frac{1}{3},-1,0}(x)+\frac{1}{2}G_{-\frac{1}{3},0,0}(x)+\frac{1}{2}G_{-\frac{1}{3},1,0}(x)\nonumber \\
 & +G_{0,-1,0}(x)+3G_{0,0,-1}(x)-\frac{3}{2}G_{0,0,0}(x)-2G_{0,-i,-1}(x)+G_{0,-i,0}(x)-2G_{0,i,-1}(x)\nonumber \\
 & +G_{0,i,0}(x)-\frac{1}{2}G_{0,1,0}(x)-2G_{1,0,-1}(x)+G_{1,0,0}(x)-\frac{3}{2}\log(2)G_{0,0}(x)+\log(2)G_{0,-i}(x)\nonumber \\
 & +\log(2)G_{0,i}(x)+\log(2)G_{1,0}(x)-\frac{2}{3}\pi^{2}G_{-3}(x)+\frac{1}{3}\pi^{2}G_{-\frac{1}{3}}(x)-\frac{3}{8}\pi^{2}G_{0}(x)\nonumber \\
 & +\frac{1}{6}\pi^{2}G_{1}(x)-\frac{1}{2}\log^{2}(2)G_{0}(x)+\frac{1}{48}\left(115\zeta(3)+8\pi^{2}\log(2)-32\pi^{2}\log(3)\right)]+{\cal O}(\epsilon^{4}),\\
 & F_{67}=\epsilon^{3}[\frac{1}{2}G_{-3,-1,0}(x)-\frac{1}{4}G_{-3,1,0}(x)-G_{-1,0,-1}(x)+\frac{1}{2}G_{-1,0,0}(x)+G_{-1,-i,-1}(x)\nonumber \\
 & -\frac{1}{2}G_{-1,-i,0}(x)+G_{-1,i,-1}(x)-\frac{1}{2}G_{-1,i,0}(x)-\frac{1}{2}G_{-\frac{1}{3},-1,0}(x)+\frac{1}{4}G_{-\frac{1}{3},0,0}(x)\nonumber \\
 & +\frac{1}{4}G_{-\frac{1}{3},1,0}(x)-\frac{1}{2}G_{0,-1,0}(x)-\frac{3}{2}G_{0,0,-1}(x)+\frac{3}{4}G_{0,0,0}(x)+2G_{0,-i,-1}(x)-G_{0,-i,0}(x)\nonumber \\
 & +2G_{0,i,-1}(x)-G_{0,i,0}(x)+\frac{1}{4}G_{0,1,0}(x)+2G_{-i,0,-1}(x)-G_{-i,0,0}(x)-2G_{-i,-i,-1}(x)\nonumber \\
 & +G_{-i,-i,0}(x)-2G_{-i,i,-1}(x)+G_{-i,i,0}(x)+2G_{i,0,-1}(x)-G_{i,0,0}(x)-2G_{i,-i,-1}(x)\nonumber \\
 & +G_{i,-i,0}(x)-2G_{i,i,-1}(x)+G_{i,i,0}(x)+G_{1,0,-1}(x)-\frac{1}{2}G_{1,0,0}(x)-G_{1,-i,-1}(x)\nonumber \\
 & +\frac{1}{2}G_{1,-i,0}(x)-G_{1,i,-1}(x)+\frac{1}{2}G_{1,i,0}(x)+\frac{1}{2}\log(2)G_{-1,0}(x)\nonumber \\
 & -\frac{1}{2}\log(2)G_{-1,-i}(x)-\frac{1}{2}\log(2)G_{-1,i}(x)+\frac{3}{4}\log(2)G_{0,0}(x)-\log(2)G_{0,-i}(x)\nonumber \\
 & -\log(2)G_{0,i}(x)-\log(2)G_{-i,0}(x)+\log(2)G_{-i,-i}(x)+\log(2)G_{-i,i}(x)-\log(2)G_{i,0}(x)\nonumber \\
 & +\log(2)G_{i,-i}(x)+\log(2)G_{i,i}(x)-\frac{1}{2}\log(2)G_{1,0}(x)+\frac{1}{2}\log(2)G_{1,-i}(x)\nonumber \\
 & +\frac{1}{2}\log(2)G_{1,i}(x)+\frac{1}{3}\pi^{2}G_{-3}(x)-\frac{5}{6}\pi^{2}G_{-1}(x)+\frac{1}{6}\pi^{2}G_{-\frac{1}{3}}(x)\nonumber \\
 & +\frac{11}{48}\pi^{2}G_{0}(x)-\frac{1}{6}\pi^{2}G_{-i}(x)-\frac{1}{6}\pi^{2}G_{i}(x)-\frac{1}{24}\pi^{2}G_{1}(x)+\frac{1}{4}\log^{2}(2)G_{-1}(x)\nonumber \\
 & +\frac{1}{2}\log^{2}(2)G_{0}(x)-\frac{1}{2}\log^{2}(2)G_{-i}(x)-\frac{1}{2}\log^{2}(2)G_{i}(x)-\frac{1}{4}\log^{2}(2)G_{1}(x)\nonumber \\
 & +\frac{1}{96}\left(-235\zeta(3)+16\log^{3}(2)+4\pi^{2}\log(2)+32\pi^{2}\log(3)\right)]+{\cal O}(\epsilon^{4}),\\
 & F_{68}={\cal O}(\epsilon^{4}),\\
 & F_{69}={\cal O}(\epsilon^{4}),\\
 & F_{70}=\epsilon^{3}[i\pi G_{-3,-1}(z)-i\pi G_{1,-1}(z)-G_{-3,-1,-1}(z)+2G_{-3,-1,1}(z)-G_{-1,0,1}(z)\nonumber \\
 & +G_{1,-1,-1}(z)-2G_{1,-1,1}(z)+G_{1,0,1}(z)-\log(2)G_{-3,-1}(z)+\log(2)G_{1,-1}(z)\nonumber \\
 & -2\text{Li}_{3}\left(\frac{4}{3}\right){}^{*}+\frac{1}{6}\pi^{2}G_{-3}(z)-\frac{1}{6}\pi^{2}G_{-1}(z)+\frac{1}{2}\log^{2}(2)G_{-3}(z)-\frac{1}{2}\log^{2}(2)G_{1}(z)\nonumber \\
 & -i\pi\log(2)G_{-3}(z)+i\pi\log(2)G_{1}(z)\nonumber \\
 & +2\left(\text{Li}_{3}\left(\frac{2}{3}\right)+\text{Li}_{3}(2)\right)-2\text{Li}_{3}\left(\frac{1}{3}\right)-2\text{Li}_{3}\left(\frac{1}{4}\right)\nonumber \\
 & -\text{Li}_{3}\left(-\frac{1}{2}\right)+2\text{Li}_{3}(-2)-\text{Li}_{2}\left(\frac{1}{3}\right)\log(4)-\text{Li}_{2}\left(-\frac{1}{2}\right)(\log(8)-i\pi)\nonumber \\
 & -4\text{Li}_{2}(2)\coth^{-1}(7)+\frac{\zeta(3)}{4}+\frac{1}{6}\log^{2}(2)(15\log(3)+\log(256))-\frac{1}{3}\log^{2}(3)\log\left(\frac{64}{3}\right)\nonumber \\
 & +\frac{1}{6}i\pi\left(\pi^{2}+6\log^{2}(3)-i\pi\log\left(\frac{128}{81}\right)-9\log(2)\log\left(\frac{9}{2}\right)\right)]+{\cal O}(\epsilon^{4}),\\
 & F_{71}=\epsilon^{2}[\frac{1}{4}\left(2H_{-1,-1}(z)+\log(2)\left(\log(2)-2H_{-1}(z)\right)\right)]\nonumber \\
 & +\epsilon^{3}[\frac{1}{24}(-36H_{-1,-1,-1}(z)-24H_{-1,1,-1}(z)+24H_{1,-1,-1}(z)\nonumber \\
 & +(-12\log(2)+24i\pi)H_{-1,-1}(z)+24\log(2)H_{-1,1}(z)-24\log(2)H_{1,-1}(z)\nonumber \\
 & -(2\pi^{2}-42\log^{2}(2)+24i\pi\log(2))H_{-1}(z)+12\log^{2}(2)H_{1}(z)\nonumber \\
 & -9\zeta(3)-18\log^{3}(2)+12i\pi\log^{2}(2)+2\pi^{2}\log(2))]+{\cal O}(\epsilon^{4}),\\
 & F_{72}=\epsilon^{3}[-2i\pi G_{-3,-1}(z)-i\pi G_{1,-1}(z)+2G_{-3,-1,-1}(z)-4G_{-3,-1,1}(z)+G_{-1,-1,-1}(z)\nonumber \\
 & +2G_{-1,0,1}(z)-G_{-1,1,-1}(z)+G_{1,-1,-1}(z)-2G_{1,-1,1}(z)+G_{1,0,1}(z)+2\log(2)G_{-3,-1}(z)\nonumber \\
 & +\frac{1}{2}\log(32)G_{-1,-1}(z)-2\log(8)G_{-1,-1}(z)-\frac{1}{4}\log(4)G_{-1,-1}(z)+3\log(2)G_{-1,-1}(z)\nonumber \\
 & +\log(2)G_{-1,1}(z)+\frac{1}{2}\log(4)G_{1,-1}(z)+4\text{Li}_{3}\left(\frac{4}{3}\right){}^{*}-\frac{1}{3}\pi^{2}G_{-3}(z)+\frac{5}{12}\pi^{2}G_{-1}(z)\nonumber \\
 & -\log^{2}(2)G_{-3}(z)-6\log^{2}(2)G_{-1}(z)-\frac{1}{2}\log^{2}(2)G_{1}(z)+2i\pi\log(2)G_{-3}(z)\nonumber \\
 & +2\log(2)\log(8)G_{-1}(z)+i\pi\log(2)G_{1}(z)+\frac{7\text{Li}_{3}(4)}{2}\nonumber \\
 & -4\text{Li}_{3}\left(\frac{2}{3}\right)+4\text{Li}_{3}\left(\frac{1}{3}\right)-8\text{Li}_{2}\left(\frac{2}{3}\right)\log(2)+5\text{Li}_{2}\left(\frac{1}{4}\right)\log(2)\nonumber \\
 & +\frac{1}{2}i\pi\left(-2\text{Li}_{2}\left(\frac{1}{4}\right)+4\log\left(\frac{4}{3}\right)\log(3)+\log(8)\log\left(\frac{81}{8}\right)\right)-\frac{49\zeta(3)}{8}\nonumber \\
 & +\frac{i\pi^{3}}{12}+\frac{21\log^{3}(2)}{2}-\frac{1}{3}\log^{2}(3)\log(576)-\log^{2}(2)\log(3)-\pi^{2}\log\left(\frac{4\sqrt[4]{2}}{\sqrt[3]{3}}\right)]+{\cal O}(\epsilon^{4}),\\
 & F_{73}=\epsilon^{2}[\frac{1}{4}H_{0}(x){}^{2}]+\epsilon^{3}[-\frac{1}{12}H_{0}(x)\left(12H_{-1,0}(x)-6H_{0,0}(x)+\pi^{2}\right)]+{\cal O}(\epsilon^{4}),\\
 & F_{74}=\epsilon^{2}[\left(-H_{0}(y)+i\pi\right){}^{2}]\nonumber \\
 & +\epsilon^{3}[\frac{1}{3}\left(H_{0}(y)-i\pi\right)\left(6H_{-1,0}(y)-12H_{0,0}(y)-6i\pi H_{-1}(y)+12i\pi H_{0}(y)+5\pi^{2}\right)]\nonumber \\
 &+{\cal O}(\epsilon^{4}),\\
 & F_{75}=\frac{1}{16}+\epsilon[-\frac{\log(2)}{4}+\frac{i\pi}{8}]+\epsilon^{2}[\frac{1}{48}\left(-7\pi^{2}+24\log^{2}(2)-24i\pi\log(2)\right)]\nonumber \\
 & +\epsilon^{3}[-\frac{\zeta(3)}{4}-\frac{i\pi^{3}}{8}-\frac{2\log^{3}(2)}{3}+i\pi\log^{2}(2)+\frac{7}{12}\pi^{2}\log(2)]+{\cal O}(\epsilon^{4}),\\
 & F_{76}=\epsilon^{2}[-H_{-1,0}(y)+i\pi H_{-1}(y)]\nonumber \\
 & +\epsilon^{3}[i\pi H_{-1,-1}(y)-2i\pi H_{-1,0}(y)-H_{-1,-1,0}(y)+2H_{-1,0,0}(y)-\frac{5}{6}\pi^{2}H_{-1}(y)]+{\cal O}(\epsilon^{4}),\\
 & F_{77}=\epsilon^{2}[-H_{0,0}(x)+H_{-1,0}(y)-i\pi H_{-1}(y)-\frac{\pi^{2}}{2}]\nonumber \\
 & +\epsilon^{3}[2H_{0,-1,0}(x)-H_{0,0,0}(x)-i\pi H_{-1,-1}(y)+2i\pi H_{-1,0}(y)+H_{-1,-1,0}(y)\nonumber \\
 & -2H_{-1,0,0}(y)+\frac{1}{6}\pi^{2}H_{0}(x)+\frac{5}{6}\pi^{2}H_{-1}(y)-4\zeta(3)]+{\cal O}(\epsilon^{4}),\\
 & F_{78}=\epsilon^{2}[-\frac{1}{2}H_{0,0}(x)-\frac{\pi^{2}}{4}]\nonumber \\
 & +\epsilon^{3}[H_{0,-1,0}(x)-\frac{1}{2}H_{0,0,0}(x)+\frac{1}{12}\pi^{2}H_{0}(x)-2\zeta(3)]+{\cal O}(\epsilon^{4}),\\
 & F_{79}=\epsilon^{3}[-\left(\pi+iH_{0}(y)\right)\left(\pi H_{-1}(y)+iH_{-1,0}(y)\right)]+{\cal O}(\epsilon^{4}),\\
 & F_{80}=\epsilon^{3}[\frac{1}{2}\left(H_{0}(y)-i\pi\right)\left(2H_{0,0}(x)-2H_{-1,0}(y)+2i\pi H_{-1}(y)+\pi^{2}\right)]+{\cal O}(\epsilon^{4}),\\
 & F_{81}=\epsilon^{2}[-\frac{1}{8}H_{0,0}(x)-\frac{\pi^{2}}{16}]\nonumber \\
 & +\epsilon^{3}[-\frac{1}{48}i(6\pi H_{0,0}(x)+12iH_{0,-1,0}(x)-6iH_{0,0,0}(x)+12i\log(2)H_{0,0}(x)\nonumber \\
 & +i\pi^{2}H_{0}(x)-24i\zeta(3)+3\pi^{3}+6i\pi^{2}\log(2))]+{\cal O}(\epsilon^{4}),\\
 & F_{82}=\epsilon^{3}[\frac{1}{2}H_{0}(x)\left(-H_{-1,0}(y)+i\pi H_{-1}(y)\right)]+{\cal O}(\epsilon^{4}),\\
 & F_{83}=\epsilon^{3}[-\frac{1}{4}H_{0}(x)\left(2H_{0,0}(x)-2H_{-1,0}(y)+2i\pi H_{-1}(y)+\pi^{2}\right)]+{\cal O}(\epsilon^{4}),\\
 & F_{84}=\epsilon^{3}[-\frac{1}{8}H_{0}(x)\left(2H_{0,0}(x)+\pi^{2}\right)]+{\cal O}(\epsilon^{4}),\\
 & F_{85}={\cal O}(\epsilon^{4}),\\
 & F_{86}=\epsilon^{3}[-\frac{21\zeta(3)}{2}+\frac{i\pi^{3}}{2}+2\pi^{2}\log(2)]+{\cal O}(\epsilon^{4}).
\end{flalign}
}

\vspace{3ex}
\bibliographystyle{JHEP}
\bibliography{references}

\providecommand{\href}[2]{#2}\begingroup\raggedright\begin{thebibliography}{100}

\bibitem{ding}
{\bf E598} Collaboration, J. J. Aubert, et~al., {\it {Experimental Observation of a Heavy Particle J}},  Phys. Rev. Lett. {\bf 33},
  (1974) 1404.

\bibitem{richter}
SLAC-SP-017 Collaboration, J.E. Augustin et~al., {\it {Discovery of a Narrow Resonance in $e^+ e^-$ Annihilation}},  Phys. Rev.  Lett. {\bf 33}, (1974) 1406.

\bibitem{Abe:2002rb}
Belle Collaboration,  K.~Abe, et~al.,
 {\it {Observation of double c anti-c production in e+ e- annihilation at $\sqrt{s}$ approximately 10.6-GeV}}, Phys. Rev.  Lett. {\bf 89}, (2002) 142001, [\href{https://arxiv.org/abs/hep-ex/0205104}{{\tt hep-ex/0205104}}].

\bibitem{Aubert:2005tj}
BaBar Collaboration,  B.~Aubert, et~al.,
  {\it {Measurement of double charmonium production in $e^+e^-$ annihilations at $\sqrt{s}=10.6$ GeV}},
  Phys. Rev. D {\bf 72}, (2005) 031101, [\href{https://arxiv.org/abs/hep-ex/0506062}{{\tt hep-ex/0506062}}].

\bibitem{NRQCD}
G. T. Bodwin, E. Braaten and G. P. Lepage, {\it {Rigorous QCD analysis of inclusive annihilation
and production of heavy quarkonium}}, Phys. Rev. D {\bf 51}, (1995) 1125 [Erratum-ibid. D {\bf 55}, (1997) 5853 ],
 [\href{http://xxx.lanl.gov/abs/hep-ph/9407339}{{\tt hep-ph/9407339}}].


\bibitem{Zhang:2005cha}
  Y. J. Zhang, Y. J. Gao and K. T. Chao,
 {\it {Next-to-leading order QCD correction to $e^+ e^- \rightarrow J/\psi + \eta_c$ at $\sqrt{s}$ = 10.6-GeV}},
Phys. Rev. Lett. {\bf 96}, (2006) 092001, [\href{https://arxiv.org/abs/hep-ph/0506076}{{\tt hep-ph/0506076}}].


\bibitem{Zhang:2006ay}
  Y. J. Zhang and K. T. Chao,
 {\it {Double charm production $e^+ e^- \rightarrow J/\psi+c+\bar{c} $ at B factories with next-to-leading order QCD correction}},
Phys. Rev.  Lett.  {\bf 98}, (2007) 092003, [\href{https://arxiv.org/abs/hep-ph/0611086}{{\tt hep-ph/0611086}}].

\bibitem{beneke:1998} M. Beneke, A. Signer, and V. A. Smirnov,
{\it{Two loop correction to the leptonic decay of quarkonium}},
 Phys. Rev.  Lett. {\bf 80}, (1998) 2535,
[\href{http://arxiv.org/abs/hep-ph/9712302}{{\tt hep-ph/9712302}}].


\bibitem{czarnecki:1998} A. Czarnecki and K. Melnikov,
{\it{Two loop QCD corrections to the heavy quark pair production cross-section in e+ e- annihilation near the threshold}},
 Phys. Rev.  Lett. {\bf 80}, (1998) 2531,
[\href{http://arxiv.org/abs/hep-ph/9712222}{{\tt hep-ph/9712222}}].


\bibitem{Beneke:2014qea}
  M. Beneke, Y. Kiyo, P. Marquard, A. Penin, J. Piclum, D. Seidel and M. Steinhauser,
 {\it {Leptonic decay of the $\Upsilon$(1$S$) meson at third order in QCD}},
 Phys. Rev.  Lett. {\bf 112}, (2014) 15, 151801,
[\href{http://arxiv.org/abs/1401.3005}{{\tt arXiv:1401.3005}}].

\bibitem{czarnecki:2001} A. Czarnecki and K. Melnikov,
{\it {Charmonium decays: J / psi ---> e+ e- and eta(c) ---> gamma gamma}},
Phys. Lett. B {\bf 519}, (2001) 212,
[\href{http://arxiv.org/abs/hep-ph/0109054}{{\tt hep-ph/0109054}}].


\bibitem{bc22} Andrei I. Onishchenko and Oleg L. Veretin, Eur. Phys. J. C {\bf 50}, (2007) 801, [\href{https://arxiv.org/abs/hep-ph/0302132}{{\tt hep-ph/0302132}}].

\bibitem{Chen:2015csa}
  L. B. Chen and C. F. Qiao,
  {\it {Two-loop QCD Corrections to $B_c$ Meson Leptonic Decays}},
  Phys. Lett. B {\bf 748}, (2015) 443,
[\href{http://arxiv.org/abs/1503.05122}{{\tt arXiv:1503.05122}}].


\bibitem{Feng:2015uha}
  F. Feng, Y. Jia and W. L. Sang,
 {\it {Can Nonrelativistic QCD Explain the $\gamma\gamma^* \to \eta_c$  Transition Form Factor Data}}.
 Phys. Rev. Lett.  {\bf 115}, (2015) 22, 222001.
[\href{http://arxiv.org/abs/1505.02665}{{\tt arXiv:1505.02665}}].

\bibitem{beneke:1997} M. Beneke and V. A. Smirnov,
{\it{Asymptotic expansion of Feynman integrals near threshold}},
 Nucl. Phys. B {\bf 522}, (1998) 321,
[\href{http://arxiv.org/abs/hep-ph/9711391}{{\tt hep-ph/9711391}}].


\bibitem{Kotikov:1990kg}
A. Kotikov, {\it {Differential equations method: New technique for massive
  Feynman diagrams calculation}},  Phys. Lett. B {\bf 254}, (1991) 158--164.

\bibitem{Kotikov:1991pm}
A. Kotikov, {\it {Differential equation method: The Calculation of N point
  Feynman diagrams}},  Phys. Lett. B {\bf 267}, (1991) 123--127.

\bibitem{Remiddi:1997ny}
E. Remiddi, {\it {Differential equations for Feynman graph amplitudes}},  em
  Nuovo Cim. A {\bf 110}, (1997) 1435--1452,
  [\href{http://xxx.lanl.gov/abs/hep-th/9711188}{{\tt hep-th/9711188}}].


\bibitem{Gehrmann:1999as}
T. Gehrmann and E. Remiddi, {\it {Differential equations for two loop four
  point functions}},  Nucl. Phys. B {\bf 580}, (2000) 485--518,
  [\href{http://arxiv.org/abs/hep-ph/9912329}{{\tt hep-ph/9912329}}].

\bibitem{Argeri:2007up}
M. Argeri and P. Mastrolia, {\it {Feynman Diagrams and Differential
  Equations}},  Int. J. Mod. Phys. A {\bf 22}, (2007) 4375--4436,
  [\href{http://arxiv.org/abs/0707.4037}{{\tt arXiv:0707.4037}}].

\bibitem{Henn:2013pwa}
J. M. Henn, {\it {Multiloop integrals in dimensional regularization made
  simple}},  Phys. Rev. Lett. {\bf 110}, (2013) 251601,
  [\href{http://arxiv.org/abs/1304.1806}{{\tt arXiv:1304.1806}}].

\bibitem{Henn:2013nsa}
J. M. Henn, A. V. Smirnov, and V.~A. Smirnov, {\it {Evaluating single-scale
  and/or non-planar diagrams by differential equations}},  JHEP {\bf
  1403}, (2014) 088, [\href{http://arxiv.org/abs/1312.2588}{{\tt
  arXiv:1312.2588}}].


\bibitem{Henn:2014qga}
J. M. Henn, {\it {Lectures on differential equations for Feynman integrals}},
  J. Phys. A {\bf 48}, (2015) 153001,
  [\href{http://arxiv.org/abs/1412.2296}{{\tt arXiv:1412.2296}}].

\bibitem{Argeri:2014qva}
M. Argeri, S. Di Vita, P. Mastrolia, E. Mirabella, J. Schlenk, et al., {\it
  {Magnus and Dyson Series for Master Integrals}},  JHEP {\bf 1403},
  (2014) 082, [\href{http://arxiv.org/abs/1401.2979}{{\tt arXiv:1401.2979}}].

\bibitem{Henn:2013woa}
J. M. Henn and V. A. Smirnov, {\it {Analytic results for two-loop master
  integrals for Bhabha scattering I}},  JHEP {\bf 1311}, (2013) 041,
  [\href{http://arxiv.org/abs/1307.4083}{{\tt arXiv:1307.4083}}].


\bibitem{Henn:2014lfa}
J. M. Henn, K. Melnikov, and V. A. Smirnov, {\it {Two-loop planar master
  integrals for the production of off-shell vector bosons in hadron
  collisions}},  JHEP {\bf 1405}, (2014) 090,
  [\href{http://arxiv.org/abs/1402.7078}{{\tt arXiv:1402.7078}}].

\bibitem{Gehrmann:2014bfa}
T. Gehrmann, A. von Manteuffel, L. Tancredi, and E. Weihs, {\it {The two-loop
  master integrals for $q\overline{q} \to VV$}},  JHEP {\bf 1406}, (2014) 032, [\href{http://arxiv.org/abs/1404.4853}{{\tt arXiv:1404.4853}}].

\bibitem{Caola:2014lpa}
F. Caola, J. M. Henn, K. Melnikov, and V. A. Smirnov, {\it {Non-planar master
  integrals for the production of two off-shell vector bosons in collisions of
  massless partons}},  JHEP {\bf 1409}, (2014) 043,
  [\href{http://arxiv.org/abs/1404.5590}{{\tt arXiv:1404.5590}}].

\bibitem{DiVita:2014pza}
S. Di Vita, P. Mastrolia, U. Schubert, and V. Yundin, {\it {Three-loop master
  integrals for ladder-box diagrams with one massive leg}},  JHEP {\bf
  09}, (2014) 148, [\href{http://arxiv.org/abs/1408.3107}{{\tt
  arXiv:1408.3107}}].

\bibitem{Bell:2014zya}
G. Bell and T. Huber, {\it {Master integrals for the two-loop penguin
  contribution in non-leptonic $B$-decays}},  JHEP {\bf 1412}, (2014) 129, [\href{http://arxiv.org/abs/1410.2804}{{\tt arXiv:1410.2804}}].

\bibitem{Huber:2015bva}
  T. Huber and S. Krankl,
  {\it {Two-loop master integrals for non-leptonic heavy-to-heavy decays}},
  JHEP {\bf 1504}, (2015) 140,
 [\href{http://arxiv.org/abs/1503.00735}{{\tt arXiv:1503.00735}}].

\bibitem{Bonciani:2015eua}
R. Bonciani, V. Del Duca, H. Frellesvig, J. M. Henn, F. Moriello, and V. A.
  Smirnov, {\it {Next-to-leading order QCD corrections to the decay width $H
  \rightarrow Z\gamma$}},  JHEP {\bf 08}, (2015) 108,
  [\href{http://arxiv.org/abs/1505.00567}{{\tt arXiv:1505.00567}}].

\bibitem{Gehrmann:2015dua}
T. Gehrmann, S. Guns, and D. Kara, {\it {The rare decay $H\to Z\gamma$ in
  perturbative QCD}},  JHEP {\bf 09}, (2015) 038,
  [\href{http://arxiv.org/abs/1505.00561}{{\tt arXiv:1505.00561}}].

\bibitem{Grozin:2015kna}
A. Grozin, J. M. Henn, G. P. Korchemsky, and P. Marquard, {\it {The three-loop
  cusp anomalous dimension in QCD and its supersymmetric extensions}}, JHEP {\bf 01}, (2016) 140, [\href{http://arxiv.org/abs/1510.07803}{{\tt
  arXiv:1510.07803}}].


\bibitem{Bonciani:2016ypc}
R. Bonciani, S. Di Vita, P. Mastrolia, and U. Schubert, {\it {Two-Loop Master
  Integrals for the mixed EW-QCD virtual corrections to Drell-Yan scattering}}, JHEP {\bf 09}, (2016) 091,
   \href{http://arxiv.org/abs/1604.08581}{{\tt arXiv:1604.08581}}.


\bibitem{Smirnov:2008iw}
A. Smirnov, {\it {Algorithm FIRE -- Feynman Integral REduction}},  JHEP
  {\bf 0810}, (2008) 107, [\href{http://arxiv.org/abs/0807.3243}{{\tt
  arXiv:0807.3243}}].

\bibitem{Smirnov:2013dia}
A. Smirnov and V. Smirnov, {\it {FIRE4, LiteRed and accompanying tools to solve integration by parts relations}},  Comput. Phys. Commun. {\bf 184}, (2013) 2820--2827, [\href{http://arxiv.org/abs/1302.5885}{{\tt arXiv:1302.5885}}].

\bibitem{Smirnov:2014hma}
A. V. Smirnov, {\it {FIRE5: a C++ implementation of Feynman Integral
  REduction}},  Comput. Phys. Commun. {\bf 189}, (2014) 182--191,
  [\href{http://arxiv.org/abs/1408.2372}{{\tt arXiv:1408.2372}}].


\bibitem{Goncharov:1998kja}
A. B. Goncharov, {\it {Multiple polylogarithms, cyclotomy and modular
  complexes}},  Math. Res. Lett. {\bf 5}, (1998) 497--516,
  [\href{http://arxiv.org/abs/1105.2076}{{\tt arXiv:1105.2076}}].


\bibitem{Chen}
K.-T. Chen, {\it {Iterated path integrals}},  Bull. Amer. Math. Soc. {\bf 83}, (1977) 831--879.

\bibitem{Remiddi:1999ew}
E. Remiddi and J. Vermaseren, {\it {Harmonic polylogarithms}}, Int. J. Mod. Phys. A {\bf 15}, (2000) 725--754,
  [\href{http://arxiv.org/abs/hep-ph/9905237}{{\tt hep-ph/9905237}}].


\bibitem{Vollinga:2004sn}
  J. Vollinga and S. Weinzierl,
  {\it {Numerical evaluation of multiple polylogarithms}},
  Comput. Phys. Commun. {\bf 167}, (2005) 177,
 [\href{http://xxx.lanl.gov/abs/hep-ph/0410259}{{\tt hep-ph/0410259}}].


\bibitem{Bauer:2000cp}
  C. W. Bauer, A. Frink and R. Kreckel,
{\it {Introduction to the GiNaC framework for symbolic computation within the C++ programming language}},
 J. Symb. Comput. {\bf 33}, (2002) 1,
 [\href{http://xxx.lanl.gov/abs/cs/0004015}{{\tt arXiv:cs/0004015}}].


\bibitem{Maitre:2005uu}
  D. Maitre,
  {\it {HPL, a mathematica implementation of the harmonic polylogarithms}},
  Comput. Phys. Commun.  {\bf 174}, (2006) 222,
[\href{http://xxx.lanl.gov/abs/hep-ph/0507152}{{\tt hep-ph/0507152}}].


\bibitem{Maitre:2007kp}
  D. Maitre,
  {\it {Extension of HPL to complex arguments}},
 Comput. Phys. Commun.  {\bf 183}, (2012) 846,
[\href{http://xxx.lanl.gov/abs/hep-ph/0703052}{{\tt hep-ph/0703052}}].

\bibitem{Frellesvig:2016ske}
  H. Frellesvig, D. Tommasini and C. Wever,
  {\it {On the reduction of generalized polylogarithms to $\text{Li}_n$ and $\text{Li}_{2,2}$ and on the evaluation thereof}},
  JHEP {\bf 1603}, (2016) 189,
 [\href{http://xxx.lanl.gov/abs/1601.02649}{{\tt arXiv:1601.02649}}].

\bibitem{Adams:2015gva}
L. Adams, C. Bogner, and S. Weinzierl,
\newblock J. Math. Phys. {\bf 56}, (2015) 072303,
[\href{http://xxx.lanl.gov/abs/1504.03255}{{\tt arXiv:1504.03255}}].


\bibitem{Adams:2015ydq}
L. Adams, C. Bogner, and S.~Weinzierl,
\newblock J. Math. Phys. {\bf 57}, (2016) 032304,
[\href{http://xxx.lanl.gov/abs/1512.05630}{{\tt arXiv:1512.05630}}].


\bibitem{Adams:2014vja}
L. Adams, C. Bogner, and S. Weinzierl,
\newblock J. Math. Phys. {\bf 55}, (2014) 102301,
[\href{http://xxx.lanl.gov/abs/1405.5640}{{\tt arXiv:1405.5640}}].

\bibitem{Remiddi:2013joa}
E. Remiddi and L. Tancredi,
\newblock Nucl. Phys. B {\bf 880}, (2014) 343,
[\href{http://xxx.lanl.gov/abs/1311.3342}{{\tt arXiv:1311.3342}}].

\bibitem{Adams:2016xah}
  L. Adams, C. Bogner, A. Schweitzer and S. Weinzierl,
  {\it {The kite integral to all orders in terms of elliptic polylogarithms}},
[\href{http://xxx.lanl.gov/abs/1607.01571}{{\tt arXiv:1607.01571}}].


\bibitem{Remiddi:2016gno}
E. Remiddi and L. Tancredi,
\newblock Nucl. Phys. B {\bf 907}, (2016) 400, 
[\href{http://xxx.lanl.gov/abs/1602.01481}{{\tt arXiv:1602.01481}}].

\bibitem{Bonciani:2016qxi}
  R. Bonciani, V. Del Duca, H. Frellesvig, J. M. Henn, F. Moriello and V. A. Smirnov,
   {\it {Two-loop planar master integrals for Higgs$\to 3$ partons with full heavy-quark mass dependence}}, JHEP {\bf 1612}, (2016) 096,
 [\href{http://arxiv.org/abs/1609.06685}{{\tt arXiv:1609.06685}}].

\bibitem{inpreparation}
L. B. Chen, Y. Liang, and C. F. Qiao, in preparation.


\bibitem{Prausa:2017ltv}
  M. Prausa,
  {\it {epsilon: A tool to find a canonical basis of master integrals}},
  [\href{http://arxiv.org/abs/1701.00725}{{\tt arXiv:1701.00725}}].

\bibitem{Gituliar:2017vzm}
  O. Gituliar and V. Magerya, {\it {Fuchsia: a tool for reducing differential equations for Feynman master integrals to epsilon form}},
[\href{https://arxiv.org/abs/1701.04269}{{\tt arXiv:1701.04269}}].

\bibitem{Czakon:2005rk}
M. Czakon, {\it {Automatized analytic continuation of Mellin-Barnes
  integrals}},  Comput. Phys. Commun. {\bf 175}, (2006) 559--571,
  [\href{http://xxx.lanl.gov/abs/hep-ph/0511200}{{\tt hep-ph/0511200}}].

\bibitem{Gluza:2007rt}
J. Gluza, K. Kajda, and T. Riemann, {\it {AMBRE: A Mathematica package for the
  construction of Mellin-Barnes representations for Feynman integrals}}, Comput. Phys. Commun. {\bf 177}, (2007) 879--893,
  [\href{http://xxx.lanl.gov/abs/0704.2423}{{\tt arXiv:0704.2423}}].

\bibitem{Gluza:2010rn}
  J. Gluza, K. Kajda, T. Riemann and V. Yundin,
 {\it {Numerical Evaluation of Tensor Feynman Integrals in Euclidean Kinematics}}, Eur. Phys. J. C {\bf 71}, (2011) 1516,
  [\href{http://arxiv.org/abs/1010.1667}{{\tt arXiv:1010.1667}}].

\bibitem{Blumlein:2014maa}
  J. Bl¨¹mlein, I. Dubovyk, J. Gluza, M. Ochman, C. G. Raab, T. Riemann and C. Schneider, {\it {Non-planar Feynman integrals, Mellin-Barnes representations, multiple sums}}, PoS LL {\bf 2014}, (2014) 052,
 [\href{http://arxiv.org/abs/1407.7832}{{\tt arXiv:1407.7832}}].


\bibitem{Ferguson:1999}
H. Ferguson, D. H. Bailey, and S. Arno, {\it { Analysis of PSLQ, an integer relation finding algorithm}},  Math. Comput. {\bf 68 }, (1999) 351.


\bibitem{Broadhurst}
D. J. Broadhurst, Z. Phys. C {\bf 47}, (1990) 115.

\bibitem{Davydychev:2003mv}
  A. I. Davydychev and M. Y. Kalmykov,
  {\it {Massive Feynman diagrams and inverse binomial sums}},
  Nucl. Phys. B {\bf 699}, (2004) 3,
 [\href{http://xxx.lanl.gov/abs/hep-th/0303162}{{\tt hep-th/0303162}}].


\bibitem{Smirnov:2013eza}
A. V. Smirnov, {\it {FIESTA 3: cluster-parallelizable multiloop numerical
  calculations in physical regions}},  Comput. Phys. Commun. {\bf 185},
  (2014) 2090--2100, [\href{http://arxiv.org/abs/1312.3186}{{\tt
  arXiv:1312.3186}}].

\bibitem{Smirnov:2015mct}
  A. V. Smirnov,
 {\it {FIESTA4: Optimized Feynman integral calculations with GPU support}},
  Comput. Phys. Commun.  {\bf 204}, (2016) 189,
[\href{http://arxiv.org/abs/1511.03614}{{\tt arXiv:1511.03614}}].

\bibitem{Borowka:2012yc}
  S. Borowka, J. Carter and G. Heinrich,
  {\it {Numerical Evaluation of Multi-Loop Integrals for Arbitrary Kinematics with SecDec 2.0}}, Comput. Phys. Commun. {\bf 184}, (2013) 396,
[\href{http://arxiv.org/abs/1204.4152}{{\tt arXiv:1204.4152}}].

\bibitem{Borowka:2015mxa}
  S. Borowka, G. Heinrich, S. P. Jones, M. Kerner, J. Schlenk and T. Zirke,
  {\it {SecDec-3.0: numerical evaluation of multi-scale integrals beyond one loop}}, Comput. Phys. Commun. {\bf 196}, (2015) 470,
[\href{http://arxiv.org/abs/1502.06595}{{\tt arXiv:1502.06595}}].

\end{thebibliography}\endgroup

\end{document}